\def\noeditingmarks{}  

\documentclass[10pt,twocolumn]{article}

\usepackage{osdi2020-submit}

\usepackage{subcaption} 
\usepackage{booktabs}  
\usepackage{dcolumn}   
\usepackage{multirow}  
\usepackage{rotating}
\usepackage{xspace}
\usepackage{hyphenat}  
\usepackage{color}
\usepackage{enumerate}
\usepackage{balance}
\usepackage{fancyvrb}
\usepackage[frozencache]{minted}
\fvset{%
fontsize=\footnotesize,
numbers=left,
xleftmargin=18pt
}
\usepackage{textcomp}
\usepackage{amsmath}
\usepackage{amssymb}
\usepackage{amsfonts}
\usepackage{wasysym}
\usepackage{lastpage}
\usepackage{tabularx}
\usepackage{pifont}
\usepackage{microtype} 
\usepackage{mathtools} 
\usepackage{mathrsfs}

\usepackage[title]{appendix}

\mathchardef\mhyphen="2D

\usepackage{url}

\usepackage{ulem}
\usepackage{array}    

\usepackage{dblfloatfix}

\usepackage[noend]{algpseudocode}
\algrenewcomment[1]{\hfill// #1}
\algtext*{Function}
\algrenewcommand{\algorithmicrequire}{\textbf{Input:}}
\algrenewcommand{\algorithmicensure}{\textbf{Output:}}
\algnewcommand{\LineComment}[1]{\State // #1}
\algrenewcommand\textproc{}

\usepackage[noeka]{mathrmletter}
\usepackage{arydshln}
\usepackage[english]{babel}
\usepackage{blindtext}

\newif\ifminted
\newif\ifintrouble
\newif\ifcuttext

\ifintrouble
\cuttexttrue
\else
\cuttextfalse
\fi

\ifx\buildminted\undefined
\else
  \mintedtrue
\fi
\ifminted
  \usepackage{minted}
\else

\usepackage{colortbl} 
\fi

\usepackage[dvipsnames]{xcolor}

\definecolor{evaldata}{HTML}{008000}
\definecolor{MPss-col}{HTML}{0E6FB0}
\newcommand{\textred}[1]{\begingroup \color{red} #1\endgroup}

\ifx\noeditingmarks\undefined
   \newcommand{\pgwrapper}[2]{\textred{#1: #2}}
   \newcommand{\pgwrapperb}[1]{\textbf{#1}}
\else
   \newcommand{\pgwrapperb}[1]{}
   \newcommand{\pgwrapper}[2]{}
\fi

\ifx\noeditingmarks\undefined

\else

\fi

\newcommand{\sys}{{Walnut}\xspace}
\newcommand{\Sys}{\sys}

\makeatletter
\renewcommand*{\@fnsymbol}[1]{\ensuremath{\ifcase#1\or \star\or \dagger\or \ddagger\or
   \mathsection\or \mathparagraph\or \|\or **\or \dagger\dagger
   \or \ddagger\ddagger \else\@ctrerr\fi}}
\makeatother

\def\hn{\usefont{OT1}{phv}{mc}{n}\selectfont}

\newcommand{\mpfont}{\hn\scriptsize}
\newcommand{\MPworker}[2]{{\color{#1}\vrule\vrule}{\marginpar{\color{#1}\mpfont #2}}}
\ifx\noeditingmarks\undefined
    \newcommand{\MP}[1]{\MPworker{red}{#1}}
    \newcommand{\MPtg}[1]{\MPworker{red}{#1}}
    \newcommand{\MPat}[1]{\MPworker{blue}{#1}}
    \newcommand{\MPss}[1]{\MPworker{MPss-col}{#1}}
    \newcommand{\MPrg}[1]{\MPworker{cyan}{#1}}
    \newcommand{\SP}[1]{\MPworker{green}{#1}}
\else
    \newcommand{\MP}[1]{}
    \newcommand{\MPtg}[1]{}
    \newcommand{\MPat}[1]{}
    \newcommand{\MPss}[1]{}
    \newcommand{\MPrg}[1]{}
    \newcommand{\SP}[1]{}
\fi

\newcommand\rmv[1]{}

\newcommand{\techReportOnly}[1]{}

\newcommand{\enc}{{\small\textsf{Enc}}\xspace}
\newcommand{\dec}{{\small\textsf{Dec}}\xspace}
\newcommand{\sign}{{\small\textsf{Sign}}\xspace}
\newcommand{\verify}{{\small\textsf{Verify}}\xspace}
\newcommand{\share}{{\small\textsf{Share}}\xspace}
\newcommand{\reconstruct}{{\small\textsf{Reconstruct}}\xspace}

\usepackage{tkz-euclide}
\usepackage{tikz}
\newcommand*\circled[1]{\tikz[baseline=(char.base)]{
            \node[shape=circle,draw,inner sep=0.5pt] (char) {#1};}}

\usepackage{mathtools}
\DeclarePairedDelimiterX{\norm}[1]{\lVert}{\rVert}{#1}

\newcommand{\cpu}{\textsc{cpu}\xspace}

\def\compactify{\itemsep0in \topsep2pt \parsep=0.00in \partopsep=0pt
\leftmargin2em}
\let\latexusecounter=\usecounter

\newenvironment{myenumerate}
  {\def\usecounter{\compactify\latexusecounter}
   \begin{enumerate}}
  {\end{enumerate}\let\usecounter=\latexusecounter}

\newenvironment{myenumerate3}
  {\def\usecounter{\itemsep1pt \topsep0pt \parsep=0ex
\partopsep=0pt \leftmargin1.3em\latexusecounter}
   \begin{enumerate}}
  {\end{enumerate}\let\usecounter=\latexusecounter}

\newenvironment{myenumerate4}
  {\def\usecounter{\itemsep=0ex \topsep1ex \parsep=1ex \partopsep=0pt
\leftmargin1.8em\latexusecounter}
   \begin{enumerate}}
  {\end{enumerate}\let\usecounter=\latexusecounter}

\newenvironment{myitemize}%
  {\begin{list}{\labelitemi}{\itemsep0in \topsep2pt \parsep0.00in
  \partopsep=0pt \leftmargin\parindent}}%
  {\end{list}}

\newenvironment{myitemize2}%
  {\begin{list}{\labelitemi}{\itemsep6pt \topsep6pt \parsep0.00in
  \partopsep=0pt \leftmargin\parindent}}%
  {\end{list}}

\newenvironment{myitemize3}%
  {\begin{list}{\labelitemi}{\itemsep3pt \topsep3pt \parsep0.00in
  \partopsep=0pt \leftmargin\parindent}}%
  {\end{list}}
\newenvironment{myitemize4}%
  {\begin{list}{\labelitemi}{\itemsep0in \topsep2pt \parsep0.00in
  \partopsep=0pt \leftmargin\parindent}}%
  {\end{list}}

\newenvironment{myitemize5}%
  {\begin{list}{\labelitemi}{\itemsep3pt \topsep3pt \parsep0.00in
  \partopsep=0pt \leftmargin1.3em}}%
  {\end{list}}

\usepackage{amsthm}

\theoremstyle{definition}
\newtheorem{definition}{Definition}[section]

\theoremstyle{lemma}

\theoremstyle{theorem}

\theoremstyle{corollary}

\theoremstyle{conjecture}

\def\emparagraph#1{\vspace{0.9mm}\noindent{\bf #1}}

\def\discretionaryslash{\discretionary{/}{}{/}}
{\catcode`\/\active
\gdef\URLprepare{\catcode`\/\active\let/\discretionaryslash
        \def~{\char`\~}}}%
\def\URL{\bgroup\URLprepare\realURL}%
\def\realURL#1{\tt #1\egroup}%

\newcommand{\func}{\mathcal{F}}

\newcommand{\adversary}{\mathcal{A}}
\newcommand{\simr}{\mathsf{Sim}}

\begin{document}

\newcommand{\supsyml}[1]{\raisebox{4pt}{\footnotesize #1}}
\newcommand{\rstar}{\supsyml{$\ast$}\xspace}
\newcommand{\rdag}{\supsyml{$\ast\ast$}\xspace}

\author{ %
        Sandy Schoettler\thanks{Student authors contributed equally to the paper.}\quad
        Andrew Thompson$^{\star}$\quad
        Rakshith Gopalakrishna$^{\star}$\quad
        Trinabh Gupta\quad \\
       [6pt]\fontsize{9.5}{11}\selectfont University of California, Santa Barbara
}

\date{}

\title{\textbf{\Sys: A low-trust trigger-action platform}}
\maketitle

\begin{abstract}
Trigger-action platforms are a new type of system that connect 
    IoT devices with web services. 
For example, the popular IFTTT platform can connect Fitbit with Google Calendar 
    to add a bedtime reminder based on 
sleep history.
However, these platforms present confidentiality and integrity risks 
    as they run on public cloud infrastructure and compute over
    sensitive user data. 
This paper describes the design, implementation, and evaluation of \sys, a
low-trust trigger-action platform that
    mimics the functionality of IFTTT, while
        ensuring confidentiality of data and correctness of
        computation, 
    at a low resource cost.
The key enabler for \sys is a new two-party secure computation protocol that 
    (i)
        efficiently performs strings substitutions, which is a common
        computation in trigger-action platform workloads,
    and 
    (ii)
    replicates computation over heterogeneous
        trusted-hardware machines from different vendors
    to ensure
    correctness of computation output as long as 
        one of the machines is not compromised.  
An evaluation of \sys demonstrates its plausible deployability and low
overhead relative to a non-secure baseline---3.6$\times$ in \cpu and
4.3$\times$ in network for all but a small percentage of programs. 

\end{abstract}

\section{Introduction} 

\label{s:intro}
\label{s:introduction}

The platforms that manage heterogeneous Internet of Things (IoT) devices
have
become mainstream~\cite{googlehome, dixon2012operating, ifttt, zapier,
microsoftpowerautomate, smartthings}.
However, these platforms 
    undermine user
confidentiality and integrity as they 
    compute over sensitive user data.
Therefore, a relevant question is: 
how can we build a system that
can manage IoT devices without incurring the security risks?

To expand on the motivation above, 
consider the case of trigger-action
platforms~\cite{ifttt, zapier, microsoftpowerautomate, integromat}.
        They connect IoT devices \emph{with} web services through simple
customizable programs.
For example, a user can
    set up 
        programs to
    sync Fitbit with iOS Health~\cite{fitbitioshealth},
    adjust home temperature when they request an Uber to take them
home~\cite{nestuber},
    tweet Instagram photos~\cite{tweetinstagram},
    turn off oven if the smoke alarm detects an emergency~\cite{smokealarm}, and
    even move money from checking to savings
        depending on daily step count~\cite{fitbiting}.

Besides providing connectivity across devices and online services, trigger-action
    platforms are easy to use.
    They run in the cloud and require no local, in-home installation or maintenance. 
    Furthermore, they allow a user to go offline indefinitely after
    setting up a program. 
    Finally, their interfaces are so simple
        that an average internet user can 
                    specify programs for them
	with only a few clicks. 
    Indeed, 
    popular trigger-action platforms like IFTTT~\cite{ifttt} 
        run tens of millions of programs daily on behalf
    of their users~\cite{iftttuse,mi2017empirical}. 

However, this connectivity and convenience comes at a cost (\S\ref{s:background}).
When users opt to run programs on these platforms, 
they must provide unobstructed access to sensitive data 
    needed by the programs (and generated or consumed by the devices or
services).
Furthermore, the users assume that the platforms
    correctly perform actions on their behalf: 
    adjust thermostat to the right temperature,
        post the intended tweet, do not tamper
            with health data, and remove the right amount of money from the
    checking account.

The security risks affect not just the users: when platform providers
operate on user data, they become liable for protecting it.
Indeed, under new privacy regulations in California and EU~\cite{gdpr, ccpa},
a business can be subject to a class action lawsuit or large fine (for example, up to 4 percent of
global annual revenue) for data breaches and misuse.

Given the popularity of trigger-action platforms, there is prior
        work targeting improved security for them~\cite{xu2019privacy,fernandes2018decentralized, fernandes2017decoupled}. 
    However, as we discuss later (\S\ref{s:relwork}), 
        these works 
            provide confidentiality or integrity exclusively,
                and suffer from a fundamental weakness that 
                    shifts security risks from the platform to the connected
                        services.

In this paper, we address these limitations and demonstrate that we do not
    need to compromise significantly on security to
    support the automation provided by trigger-action platforms. 
We present \sys, a trigger-action platform that provides the same
functionality as IFTTT, but with much stronger
confidentiality and integrity protection. Furthermore, this added protection
does not affect the security at the connected services, and comes at a modest 
resource cost.

To begin with, \sys splits
    the platform into two parts that run 
    in separate administrative domains (Microsoft Azure and IBM Cloud)
    such that no part can access sensitive data~(\S\ref{s:arch}).
To obtain such a split,
    \sys employs a two-party secure computation protocol (2PC)
    called Yao's garbled circuits~\cite{yao1982protocols, zahur2015obliv},
    which allows two parties to perform arbitrary computation on
    their secret inputs without revealing their inputs to each other.
Yao's protocol is powerful as it supports 
    arbitrary computation but it is also expensive in terms of 
    \cpu time and network transfers.

\Sys makes two observations. First, typical programs on trigger-action
    platforms
    perform simple computations: 
    they either pass a sensitive string from one device or service to another without
modification,
or substitute strings from a dictionary of sensitive strings 
    into a user-supplied template string.
Second, the names of the keys in the dictionary are publicly known. 
Based on these observations, \sys designs a custom two-party secure
    computation protocol for the typical trigger-action platform programs.
\Sys's protocol leaks coarse-grained information such 
    as the approximate positions in
    the template string where the strings
    from the dictionary get substituted.
However, it
significantly reduces overhead
relative to Yao's protocol; in particular, it eliminates network 
overhead between the two parties in the protocol (\S\ref{s:design-passive}).

The 2PC protocol described so far
    allows \sys's platform to compute over
    sensitive data but does not prevent the platform from 
    deviating from the protocol and 
    producing arbitrary program output.
To efficiently add integrity guarantees, \sys
    extends the protocol to use 
    trusted-hardware
        machines such as those with Intel SGX~\cite{costan2016intel} and
    AMD SEV~\cite{kaplan2016amd} capabilities.
These machines provide a trusted execution environment (TEE)
    that can load and execute a computation
such that an external entity 
    cannot tamper with its execution.
However, a limitation of using TEEs is that 
    they introduce     
    a large trusted computing base (TCB), consisting of both a software
component and 
a hardware component.

The software part can be moved out of the TCB
    by formally verifying 
        that the software meets its required specification~\cite{bond2017vale,
            hawblitzel2014ironclad,chen2015using, ferraiuolo2017komodo,
            fonseca2017empirical}.
However, the hardware itself (the root of trust)
can be tampered, for example, through physical attacks~\cite{genkin2015get, 
    genkin2015stealing, kocher1999differential, costan2016intel}.

\Sys 
    addresses this limitation
    by employing \emph{multiple,
heterogeneous TEEs}
from different vendors.
In particular, it runs
    a copy of the 2PC protocol described above over 
    three pairs of TEEs, where 
    each counterpart in a pair
    resides in one of the two domains (for example, Microsoft Azure or IBM
Cloud).
\Sys's protocol does not guarantee an output; however, 
    if it does produce an output, then the output is correct
    as long
    as one of the TEEs is not compromised (\S\ref{s:design-active}). 
Recently, Chandran et al.~\cite{CCOV19} explored this idea of multiple TEEs 
    to obtain theoretical feasibility results in cryptography;  
however, this idea has not been explored in
practical
systems.

Besides using the two techniques above, 
    \sys's 2PC protocol composes cryptographic primitives such as end-to-end encryption, digital
signatures, and hash-chaining of OAuth authentication tokens. The latter, in particular, 
    allows a user of \sys to go offline indefinitely after
    uploading a program to run on the \sys platform (\S\ref{s:refreshtokens}).

We have implemented (\S\ref{s:impl})
    and evaluated (\S\ref{s:eval})
        a prototype of \sys. 
Our prototype
 runs
on Microsoft Azure and IBM Cloud. 
For sparsely-used programs 
    for which \sys resorts to
using Yao's protocol, it incurs a high \cpu and network
overhead: 54$\times$ for \cpu and 1883$\times$ for the network relative to a
non-secure baseline that mimics IFTTT.
However, for typical 
programs that run on IFTTT (a manual inspection of 
        50 most popular programs per the dataset
        of Mi et al.~\cite{mi2017empirical} reveals
        that 98\% of them fall in this category),
\sys incurs 3.6$\times$ \cpu and 
4.3$\times$ network overhead
relative to the non-secure baseline.
    We also implemented 30 programs from IFTTT
(15 from the set of 50 most popular ones per Mi et
al.~\cite{mi2017empirical}, and 15 selected randomly) 
        on \sys, demonstrating
that \sys is plausibly deployable.

\sys's limitations include 
    coarse-grained leakage
    from its customized 
    2PC protocol.
Nevertheless,
we believe \sys is useful: 
    it provides a benefit to both
users and platform providers, as they can enjoy trigger-action automation with
significant improvement in data confidentiality and integrity, at low resource
cost.

\section{Overview of IFTTT (IF-This-Then-That)}
\label{s:background}

\Sys aims to mimic IFTTT's functionality 
    as IFTTT is a 
    popular
    trigger-action platform~\cite{ur2016trigger, mi2017empirical,
        surbatovich2017some, fernandes2017decoupled, baruah2018two}. 
This section gives an
  overview of IFTTT and describes the
  confidentiality and integrity risks it presents.

\begin{figure}[t!]
\centerline{\includegraphics[width=3in]{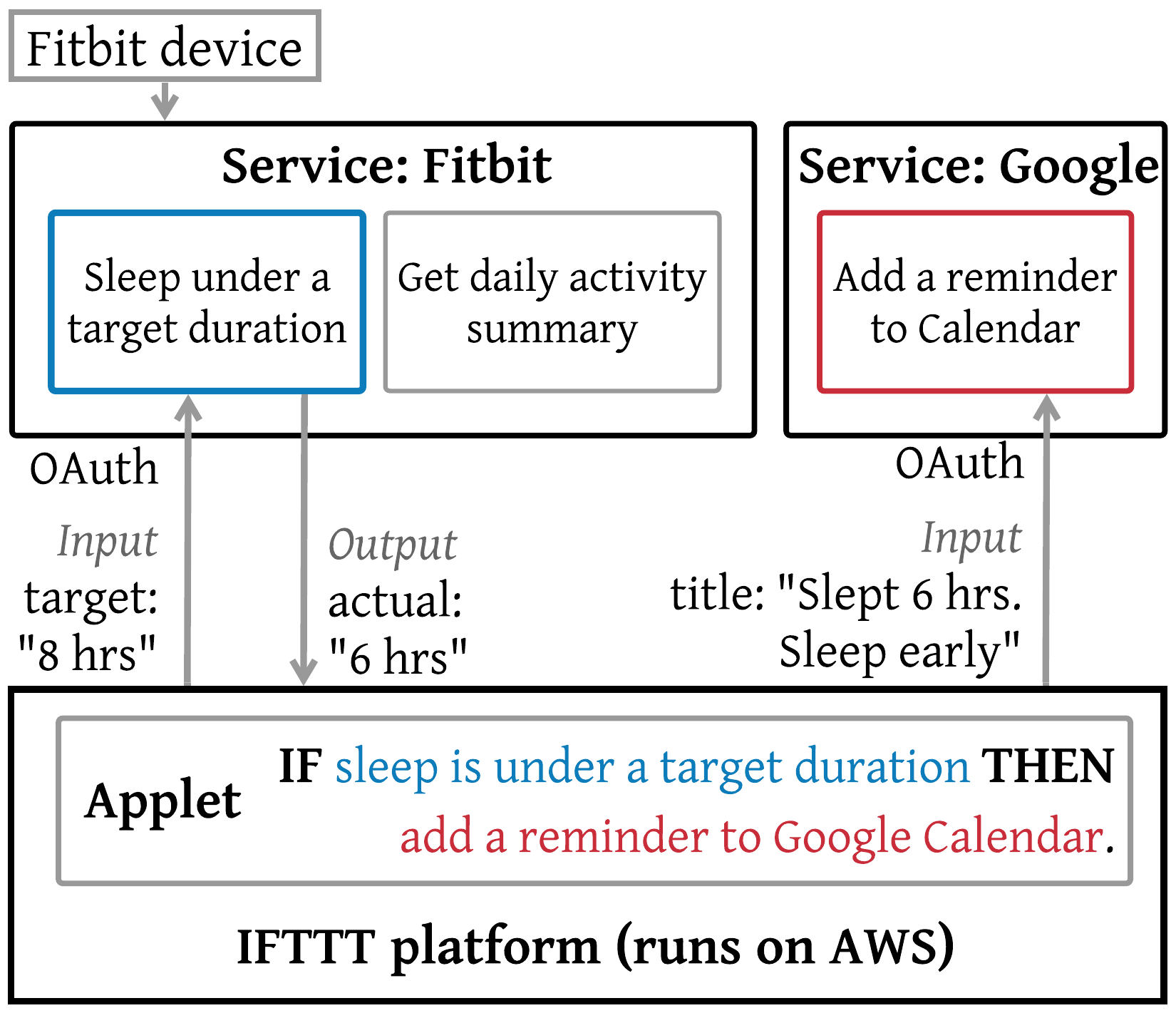}}
\caption{%
High-level architecture of IFTTT~\cite{ifttt} in the context of a program
(applet) that connects Fitbit service to Google calendar service. 
	}
\label{f:ifttt-arch}
\label{fig:ifttt-arch}
\end{figure}

Figure~\ref{f:ifttt-arch} depicts 
    IFTTT's architecture, which consists of three key components: 
        \emph{services}, \emph{platform}, and \emph{applets}.

Services are either
    IoT device management services such as the Fitbit service 
    or online-only web-services such as Google calendar. 
These services 
    expose user information and provide an ability to update that information
        through HTTP-based RESTful APIs.
For example, the Fitbit service exposes an API that returns
    the sleep duration of a user. 
    Similarly, the Google calendar service exposes
    an API to add a reminder to a user's calendar.

The IFTTT platform runs in the public cloud (AWS currently). 
    It 
        aggregates APIs from 
        all services and presents them to applets. 
    It is also responsible for executing these applets. 

Applets are programs that stitch together two service APIs. 
    They are the core abstraction that IFTTT exposes.
For example, the ``When you sleep less than the desired amount, add a reminder on
your calendar to go to bed early tomorrow'' applet connects the sleep duration
API from Fitbit with the add reminder API from Google calendar. 
An applet contains a TypeScript code that takes the output of 
    the first API and manipulates it before feeding
        into the second API.
For instance, 
    the applet depicted in Figure~\ref{f:ifttt-arch} takes 
    the user's sleep duration string ``6 hrs'' returned by the Fitbit 
        API and adds it to the 
            title of the reminder.
An applet also contains OAuth authentication 
    tokens~\cite{hardt2012oauth} that allow the 
    platform to access the APIs on user behalf.

\emparagraph{Confidentiality and integrity risks.}    
IFTTT 
    acts as a delegatee and 
    runs applets on user behalf; this  
    arrangement
    naturally creates both confidentiality and 
    integrity risks. 
In particular, when a user creates an applet and loads it to the IFTTT
    platform, or when the IFTTT platform
    executes the applet, the platform learns
    sensitive information such as the APIs the user intends to use, the
parameters the user wants to send to the APIs, and the information returned
by the APIs. 
Similarly, the IFTTT platform creates integrity risks because it may 
    not run an applet as intended. 
    For instance, it may set an incorrect time for reminder 
        in the Fitbit and Google calendar example discussed above. 
        In other examples, it may cause more damage:
        turn oven on instead of off, move an incorrect amount of money from
    a user's checking to savings account, drop malware into user's dropbox
        account, etc.

The confidentiality and integrity risks described above are
significant; however, the current design of the IFTTT platform and the 
    connected services 
    exacerbates them.
The connected services give out 
    OAuth access tokens 
        with broad scope, 
        so that
            a user
            can authenticate with a service once and use the token
                across applets that call different APIs at the service~\cite{fernandes2016security,
    fernandes2018decentralized,fernandes2017decoupled}.
Thus, if the platform misbehaves, 
    it can create malicious applets 
        using the (broad) OAuth tokens.

Thus, our goal is to 
        design mechanisms that address these risks, that is, enable the IFTTT platform 
            provider to keep
           user data confidential while operating on it as intended by the user, at low
    resource costs (CPU, network, etc.).
We are not saying that the IFTTT platform provider is inherently untrustworthy.
    However, the platform provider runs the platform on a public cloud. 
            Besides, there are many cases of both 
                internal and external adversaries, such as rogue system
        administrators and hackers,
            that can get into cloud infrastructure and 
            steal and tamper with user information~\cite{roguesteffan,
roguegoogle, roguetechspot, cloudhopper, cyber20}.
    On top of that, the 
        platform provider is incentivized to
        reduce risk for user data due to the risk of significant fines and
lawsuits under modern privacy regulation~\cite{ccpa, gdpr}.

\section{Overview of \sys}
\label{s:overview}

\subsection{Architecture}
\label{s:arch}

\begin{figure}[t]
\centerline{\includegraphics[width=3.25in]{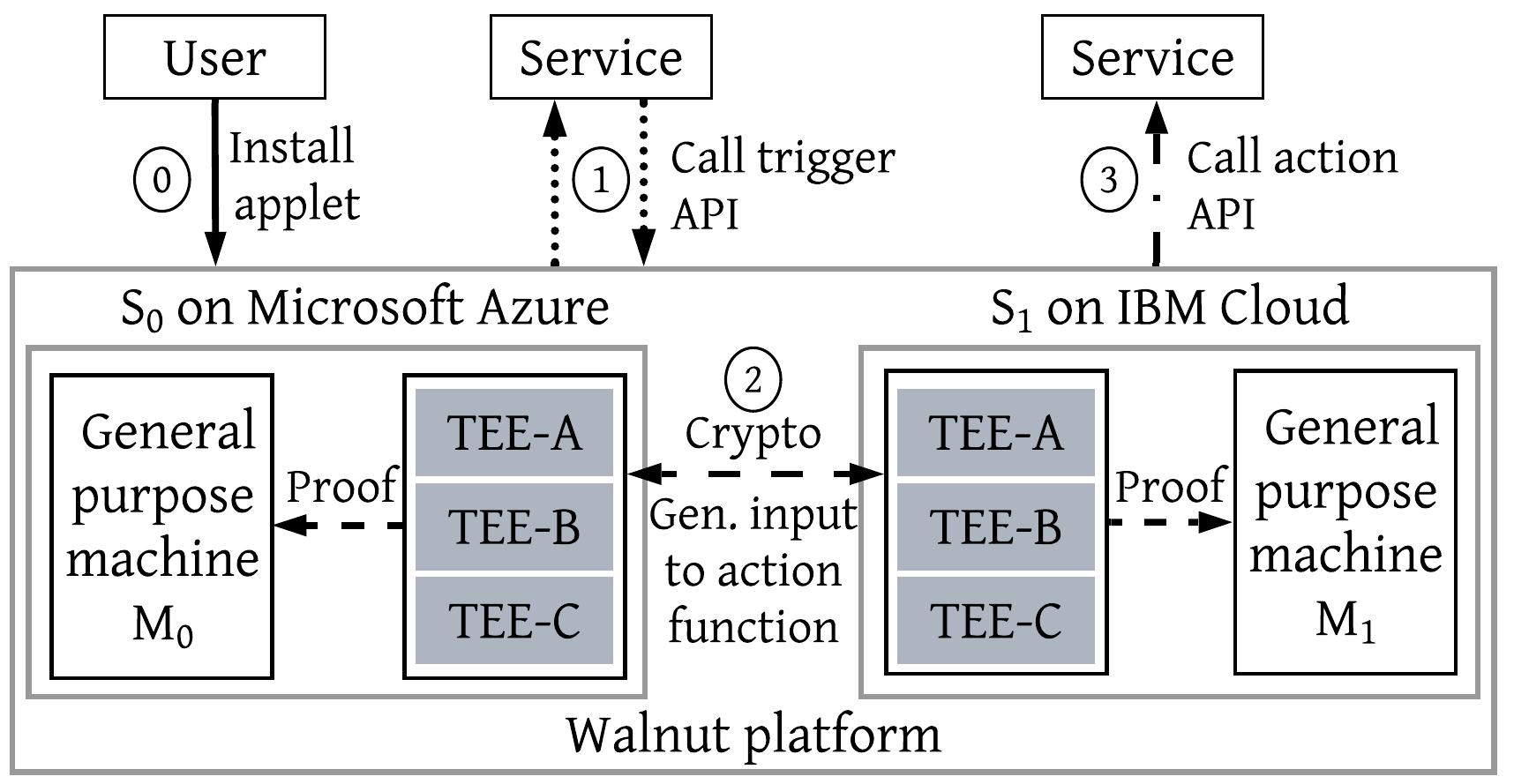}}
\caption{%
\Sys's architecture. 
Solid, dotted, dashed, and dashed-with-dots arrows
respectively depict computation performed during setup, trigger-polling,
action-generation, and action-execution phases of \sys's protocol.
}
\label{f:forttt-arch}
\label{fig:forttt-arch}
\end{figure}

Figure~\ref{f:forttt-arch} shows \sys's architecture.
\Sys has the same key components as IFTTT, namely the 
    platform, services, and applets, 
but makes modifications, particularly, 
    to the platform and the services.

\Sys's 
platform
    runs over two servers, $S_0$ and $S_1$, in separate administrative
        domains such as Microsoft Azure and IBM Cloud.
Each server is logically centralized but physically distributed; 
    it contains trusted-hardware machines from different vendors
(denoted as ``TEE-A'', ``TEE-B'', and ``TEE-C'' in Figure~\ref{f:forttt-arch})
    and general-purpose compute machines (labeled collectively as
$M_0$ and $M_1$ 
in Figure~\ref{f:forttt-arch}).

The 
    services 
    implement the HTTP APIs. Today, they 
                send or receive plaintext data (over secure channels) 
        to or from the platform. 
    Therefore, some change is necessary to support a secure
        \sys platform. \Sys expects
    these services to implement logic 
        for basic cryptographic primitives such
            as encryption, signatures, and secret-sharing (described below), 
        and to be aware of and hold public keys for \sys's two servers and their TEEs.
        This logic
    is agnostic of the API implementation, and 
        thus can be implemented once and distributed
            to all the services (for example, by the \sys 
                platform provider).

In \sys, as in IFTTT, the applet is the core programming
  abstraction. 
    \Sys does not make any change to this core abstraction.
    Each applet consists of URLs for two APIs: a \emph{trigger} API
        and an \emph{action} API. We denote
    the inputs to the APIs as \emph{trigInp} and \emph{actInp}, respectively,
        and the output from the trigger API as \emph{trigOut} (the action API
    does not produce an output).
    Each of trigInp, trigOut, and actInp is a set of 
        key-value pairs. 
    \emph{filterCode} is the code 
        component of the applet that 
        produces actInp based on trigOut.

\sys's protocol to delegate execution of an applet
    to \sys's platform 
        has four phases: \emph{setup},
            \emph{trigger-polling},
            \emph{action-generation}, and
            \emph{action-execution}.

\begin{myitemize3}

\item {\bf Setup.} 
    This phase (marked \circled{0}
        in Figure~\ref{f:forttt-arch}) runs once per applet.
    During this phase, the user creates the applet App 
        and gives its parts 
            to the two servers: part $\textrm{App}_0$ to $S_0$ and
        part $\textrm{App}_1$ to $S_1$.
   Each part contains signed ciphertexts or \emph{secret-shares} of the 
            sensitive applet data.  
         Secret-sharing refers to the
        technique 
            of splitting a secret $s$ 
                into two strings, $sh_0^{(s)}$ and $sh_1^{(s)}$, such that 
                each string appears uniformly random, but 
                    $sh_0^{(e)} \oplus sh_1^{(e)} = e$.

\item {\bf Trigger-polling.}
As in IFTTT, this phase (marked \circled{1} in Figure~\ref{f:forttt-arch})
    runs periodically (every 15 minutes) or in response to a
    real-time notification from the service implementing the trigger
API.
In this phase, 
server $S_0$ calls the trigger API in the applet. 
    In return, server $S_b$ for each $b \in \{0, 1\}$ gets a signed secret-share of 
	trigOut.

\item {\bf Action-generation.}
This phase (marked \circled{2} in Figure~\ref{f:forttt-arch}) 
    follows the trigger-polling phase.
In this phase, each \sys server 
    uses its outputs from the two phases above, 
    and runs
    a cryptographic protocol with the other server to obtain
       its secret-share of the input to the action API. 
    That is, $S_b$ obtains $sh_b^{(actInp)}$.
    Server $S_b$
    also generates a proof that it computed $sh_b^{(actInp)}$ correctly.
    This phase uses the TEE machines at the two servers.

\item {\bf Action-execution.}
Finally, in the action-execution phase (marked \circled{3} in
Figure~\ref{f:forttt-arch}), the two servers
    send their output from the action-generation phase 
        to the service implementing the action API,
    who performs checks on the proofs, merges the 
    shares of the API input, and executes
	the API.

\end{myitemize3}

\subsection{Threat model and security definitions}
\label{s:threat-model}
\label{s:security-definition}
\label{s:security-definitions}

We consider two settings for \sys, \emph{passive} and \emph{active}, depending on the
power of the adversary. In the
    passive 
    setting, 
        the adversary is honest-but-curious, meaning that it follows the
protocol description but tries to infer 
    sensitive user data by inspecting protocol messages.
In this setting, \sys uses only the general-purpose machines at $S_b$, which
    are collectively denoted as $M_b$ for $b \in \{0, 1\}$. 
    The adversary 
can compromise one of $M_b$ for
some $b \in \{0, 1\}$.

In the active setting, the adversary is malicious and can 
    arbitrarily deviate from the protocol.
In this setting, each server $S_b$
uses three types of TEEs (made by three different vendors) besides the
general-purpose machines;  
we denote
$S_b$'s three types of TEEs by $T_b^{(i)}$ for $i \in \{0, 1, 2\}$.  
The adversary can compromise one of $M_b$ for $b \in \{0, 1\}$, 
    and at most two of
$T_b^{(0)}, T_b^{(1)}$ and $T_b^{(2)}$. 

We define security notions for the two settings separately.

\vspace{-1mm}
\begin{definition}[Passive security]
\label{def:passive}
A passive \sys scheme consisting of setup, trigger-polling, action-generation, and
action-execution phases is said to be $\mathcal{L}$-secure if for any honest-but-curious
(passive) probabilistic polynomial time (PPT) adversary $\adversary$ corrupting
$M_b$ for some $b \in \{0,1\}$, for a leakage function $\mathcal{L}$, for every large enough security parameter $\lambda$, 
there exists a PPT simulator $\simr_{\mathcal{L}}$ 
        with access to the leakage function, such that 
        the output distribution of the simulator is computationally
            indistinguishable from the adversary's view in the real protocol.
\end{definition}
\vspace{-1mm}

\vspace{-1mm}
\begin{definition}[Active security]
\label{def:active}
An active \sys scheme consisting of setup, trigger-polling, action-generation, and
action-execution phases is said to be $\mathcal{L}$-secure with abort if, for any
malicious (active) probabilistic polynomial time (PPT) adversary
$\adversary$ corrupting $M_b, T_b^{(i)}, T_{b}^{(j)}$ for some $b \in \{0,1\}$
and
$i,j
\in \{0, 1, 2\}$, with black-box access to TEE $T_b^{(k)}$ implementing
functionality $\func_b^{(k)}$ for $k \in \{0,1,2\}$ and $k \neq
i,k \neq j$, for a leakage function $\mathcal{L}$, for every large enough security parameter $\lambda$, 
there exists a
PPT simulator $\simr_{\mathcal{L}}$ 
        with access to the leakage function, such that 
        the output distribution of the simulator is computationally
            indistinguishable from the adversary's view in the real protocol.

\end{definition}
\vspace{-1mm}

Essentially, the definitions say that 
    the view of the adversary in an
    execution of \sys's protocol can be simulated by 
    knowing just the inputs to the simulator, which is defined by the leakage
        function. Furthermore, 
     the protocol aborts if the adversary in the active setting deviates from
    the protocol.

    For \sys, the leakage function is defined as
$\mathcal{L}(App)$, which equals $\{\mathbb{K}(App), \mathbb{P}(App), \mathbb{F}(App)\}$.
$\mathbb{K}(App)$ equals $\{K_1, K_2, \dots, K_m\}$, where
        where $m$ is the number of key-value pairs in the actInp parameter to
        action API, and
            $K_i = \{key_1, key_2, \dots key_{k_i}\}$ contains the key names of
key-value pairs in trigOut that go into computing the $i$-th key-value pair of
actInp. 
$\mathbb{P}(App)$ equals $\{pos_1, pos_2, \dots, pos_m\}$, where 
    $pos_i = \{index_1, index_2, \dots, index_{k_i}\}$ is the set of
    positions
in the $i$-th key-value pair in actInp where the key-value pairs in trigOut
    get inserted.
$\mathbb{F}(App)$ is the filterCode for the applet and the URLs of trigger and
action APIs.

\subsection{Key techniques}
\label{s:keytechniques}

\Sys's protocol consisting of the four phases (setup, trigger-polling,
action-generation, action-execution) and meeting the security
definitions stated above uses three key techniques.

\emparagraph{Tailored 2PC for IFTTT-like applets.}
    Walnut's protocol, in the action-generation phase, uses
        a tailored two-party secure computation (2PC)
        cryptographic sub-protocol
        to enable its servers to obtain secret-shares
        of actInp.
   Walnut's 2PC protocol 
        is tailored
        to the common filterCode that generates 
            actInp fields via string substitutions, for instance, 
        by substituting the value of duration in the template string
                ``Slept \{\{duration\}\}. Sleep early.''
    Although \sys's protocol leaks a small amount of information, captured in the
        leakage function above, 
        it is significantly more efficient than a general-purpose 2PC
            protocol~\cite{yao1982protocols} (\S\ref{s:design-passive}).

\emparagraph{Proof replication across heterogeneous TEEs.}
    \Sys, during the action-generation phase, uses trusted hardware machines 
        to generate 
        hardware attestations (which we call proofs for simplicity) 
        that the code to generate action
            API input is executed inside a trusted execution environment (TEE). 
    However, relative to a solution that would use a single TEE for proof generation, 
        \sys replicates proof generation over
            heterogeneous TEEs such that 
            as long as \emph{one} TEE works correctly, 
            an adversary cannot pass a proof check and result in an incorrect
            invocation of the action API (\S\ref{s:design-active}). 

\emparagraph{Chaining of OAuth tokens.}
    \Sys, as in IFTTT, ensures that a user can go offline after executing 
        the setup phase. 
    To enable this behavior, \sys creates
        signed tuples of 
        the expirable OAuth token supplied by the user during the setup phase 
            and fresh OAuth tokens generated by the services 
	(\S\ref{s:refresh-tokens}).

\begin{figure*}[t]
\hrule
\medskip
{ 

\begin{center}
\textbf{\Sys's protocol for the passive security setting}
\vspace{-2mm}
\end{center}

\begin{myitemize5}

\item This protocol has four parties: a user, \sys's platform consisting of
    machines $M_0$ and $M_1$ in separate administrative domains, a trigger
service $TS$, and an action service $AS$. The protocol allows the user to install an
applet and the platform to execute it.

\item The protocol assumes that each entity has a (public, private) key pair
    $(pk_p, sk_p)$, for $p \in \{User, M_0, M_1, TS, AS\}$. 

\end{myitemize5}

\begin{center}
\vspace{-2mm}
\textbf{\textit{Setup phase}}
\vspace{-3mm}
\end{center}

        \begin{myenumerate4}

        \item (Encrypt OAuth tokens) 
            \label{e:encryptoauthtokens}
            The user receives OAuth access tokens
            $at\mhyphen TS$ and $at\mhyphen AS$ from $TS$ and $AS$, respectively. The user encrypts these
                tokens, that is, computes $C^{at\mhyphen TS} \gets \enc(pk_{TS},
at\mhyphen TS)$ and
            $C^{at\mhyphen AS} \gets \enc(pk_{AS}, at \mhyphen AS)$.  

        \item (Encrypt trigInp)
            \label{e:encrypttriginp}
            The user encrypts the trigInp input parameter,
            that is, computes $C^{trigInp} \gets
                \enc(pk_{TS}, trigInp)$.

        \item (Secret-share actInp) 
            \label{e:secretshareafCT}
            For each key-value pair $(k, v)$ in install-time value of actInp, the user does the
                    following:

            \begin{myenumerate3}
            \item Splits $v$ into blocks ($t_1, \ldots, t_{\ell}$) at spaces and
                and punctuation marks. For instance, splits
                    ``Slept \{\{duration\}\}. Sleep early'' 
        into (``Slept'', `` '', 
        ``\{\{duration\}\}'', ``.'', `` '', ``Sleep'', `` '', ``early'').

            \item 
                \label{e:secretshareeachblock}
                For each $t_i$, if $t_i$ is like ``\{\{key-name\}\}'', 
                sets $sh_b^{(t_i)} = t_i$. Else, secret-shares 
                    $t_i$ into $sh_0^{(t_i)}$ 
                            and $sh_1^{(t_i)}$, s.t.
                    $sh_0^{(t_i)} \oplus 
                            sh_1^{(t_i)} = t_i$. 

            \item Sets $sh_b^{(actInp)} [k] =  (sh_b^{(t_1)}, \ldots, sh_b^{(t_{\ell})})$.

            \end{myenumerate3}

        \item 
        \label{e:createapplet}
        The user sends $App_b = \{T, A_b, fc\}$ to $M_b$, where
                $T$ equals $(\textrm{triggerEndpoint}, C^{at \mhyphen TS}, C^{trigInp})$,  
                $A_b$ equals $(\textrm{actionEndpoint}, C^{at \mhyphen AS},
                    sh_b^{(actInp)})$, 
                and $fc$ is the applet filterCode.
        \end{myenumerate4}

\begin{center}
\vspace{-2mm}
\textbf{\textit{Trigger-polling phase}}
\vspace{-3mm}
\end{center}

        \begin{myenumerate4}
        \setcounter{enumi}{4}
        \item (Call trigger API) 
        \label{e:calltriggerfunction}
        Machine $M_0$ sends $T$ to the trigger service $TS$ (whose name is in
            triggerEndpoint URL in $T$).

        \item (Generate shares of trigOut) 
        \label{e:generatetriggeringredients}
        $TS$ decrypts and checks the
                OAuth token, executes the trigger API, and generates
                trigOut. Then, for each $(k, v)$ in
                trigOut, $TS$ generates $sh_b^{(trigOut)}[k]$ following
                step~\ref{e:secretshareafCT} above.

        \item $TS$ sends $sh_b^{(trigOut)}$ to $M_b$.  
              
        \end{myenumerate4}

\begin{center}
\vspace{-4mm}
\textbf{\textit{Action-generation phase}}
\vspace{-3mm}
\end{center}

        \begin{myenumerate4}
        \setcounter{enumi}{7}
         \item 
            \label{e:generateactionfields}
            (Generates shares of runtime value of actInp) $M_b$ generates $sh_b^{(actInp)} \leftarrow
\textrm{GenerateAI}(sh_b^{(trigOut)}$, $sh_b^{(actInp)}$, $fc)$. GenerateAI
further
    calls either Yao's 2PC protocol or \sys's string
        substitution function
(Figure~\ref{f:forttt-action-field-generation-code}) depending on filterCode.
        \end{myenumerate4}

\begin{center}
\vspace{-2mm}
\textbf{\textit{Action-execution phase}}
\vspace{-3mm}
\end{center}

    \begin{myenumerate4}
        \setcounter{enumi}{8}
        \item 
        \label{e:executeaction}
            $M_b$ sends 
            $(\textrm{actionEndpoint}, C^{at \mhyphen AS}, sh_b^{(actInp)})$ to $AS$, 
        who 
                reconstructs actInp, 
            validates $at\mhyphen AS$, 
                and 
executes the action API.

    \end{myenumerate4}

}
\hrule
\caption{%
\Sys's protocol for the passive setting to execute an applet. 
    This protocol does not include concatenation and padding of 
        blocks, which is described separately in text. 
This protocol further calls Yao's 
protocol~\cite{yao1982protocols} or the
function in Figure~\ref{f:forttt-action-field-generation-code}.}
\label{fig:forttt-passive-protocol}
\label{f:forttt-passive-protocol}
\end{figure*}

Besides these three techniques, \sys uses standard cryptographic primitives such
as encryption, hashing, and signatures.
The next three sections (\S\ref{s:design-passive}--\S\ref{s:refresh-tokens}) dive into the details of \sys's protocol.
We begin by making a simplifying assumption that the OAuth tokens do not expire.
Under this assumption, we first present \sys's design that meets the
security definition for the passive setting described above
(\S\ref{s:design-passive}). 
    This design does not require the TEE machines.
We then extend the design and present a protocol that meets the security
definition for the active setting (\S\ref{s:design-active}). 
Finally, we relax the simplifying assumption on OAuth token expiry
(\S\ref{s:refresh-tokens}).

\section{Design for passive security}
\label{s:designpassive}
\label{s:design-passive}

\begin{figure}[t]
{\small

\begin{minted}{python}
# 'actInp_shr' is a list of block shares from user.
# shr(pad(str)) denotes a share of a padded block.
actInp_shr = [{"type": TEXT, "value": shr(
        pad("Slept") + pad(" "))}, 
        {"type": FIELD, "key-name": "duration"},
        {"type": TEXT, "value": shr(pad(".") + pad(" ") 
        + pad("Sleep") + pad(" ") + pad("early"))}]

# Input 'trigOut_shr' from the trigger service.
# Func. maps trigOut key-names to shares of actInp.
def string_sub(trigOut_shr, actInp_shr):
    output = []
    for part in actInp_shr:
        if part.type == FIELD: # lookup key-name
            output += trigOut_shr[part.key-name]
        else:
            output += part.value
    
    return "".join(output) # concatenate the blocks 
\end{minted}
}
\caption{\Sys's custom function for string substitution. This function assumes
    that secret-shares of block of strings are both padded and concatenated
(explained further in text).}

\label{f:forttt-action-field-generation-code}
\label{fig:forttt-action-field-generation-code}
\end{figure}

Figure~\ref{fig:forttt-passive-protocol} shows \sys's protocol for the passive security setting 
    discussed in the previous section (\S\ref{s:security-definitions}). 
This protocol is optimized, in terms of performance, for the common types of
    applets on IFTTT that
    perform simple substitutions: set the value of a key-value pair in
    actInp by
        replacing parts of a template string with
        values from trigOut.
We briefly describe each phase of the protocol below; the protocol figure
contains precise details.

\emparagraph{Setup phase.}
    First, the user
    obtains OAuth tokens for the two services and encrypts 
        them
        with their public keys (step~\ref{e:encryptoauthtokens} 
            in Figure~\ref{f:forttt-passive-protocol}).
    Second, the user encrypts
        trigInp as a single blob 
    with the trigger service's public key (step~\ref{e:encrypttriginp} in
Figure~\ref{f:forttt-passive-protocol}).
    Third, the user splits
    the setup-time values in actInp into \emph{blocks} and secret-shares each
block 
(step~\ref{e:secretshareeachblock} in Figure~\ref{f:forttt-passive-protocol}).

    \emparagraph{Trigger-polling.}
    In this phase, 
        $M_0$ sends encrypted blob of trigInp to the
        trigger service implementing the trigger API. 
    The
    trigger service generates secret-shares of trigOut 
        in the same way as the user generates shares of actInp in the setup step
(step~\ref{e:generatetriggeringredients} in
Figure~\ref{f:forttt-passive-protocol}).
    For instance, 
    if one of the trigOut key-value pairs is `duration': `6 hrs',
        then the trigger service partitions the value `6 hrs' into blocks
        (``6'', `` '', ``hrs'') and secret-shares them. 

    \emparagraph{Action-generation.}
    $M_0$ and $M_1$ call Yao's garbled circuits
    protocol to run the applets's filterCode
    over shares of blocks from trigOut and (setup-time value
        of) actInp; the call to Yao happens inside the GenerateAI function in
step~\ref{e:generateactionfields} in Figure~\ref{f:forttt-passive-protocol}. 
        Yao's protocol is general-purpose and 
    can handle any arbitrary filterCode. However, 
    when the filterCode involves simple string substitutions, 
    $M_0$ and $M_1$ locally run the function \texttt{string\_sub} shown in 
    Figure~\ref{fig:forttt-action-field-generation-code}. 
   The \texttt{string\_sub} procedure returns a 
        single block 
        for each key-value pair
    of actInp. 

    \emparagraph{Action-execution.}
    $M_0$ and $M_1$ send their outputs from action-generation phase 
    along with OAuth tokens to the action service, who
    reconstructs actInp, 
    checks the OAuth tokens, and executes the action API.

\emparagraph{Concatenation and padding.}
One issue with the protocol described above is that
    it reveals to \sys the length of each block in the strings in trigOut and
    actInp.
\sys mitigates this leak by concatenation and padding.
    First, both the user and the trigger service apply
    padding to each block separately. 
        This padding is configurable: each block 
            can be padded to a pre-determined maximum size, 
        or each block can be padded to the next higher size
            from a set of allowed pre-determined sizes (for example, multiples
                of five, or powers of two).
    Second, the user and the trigger service
        concatenate padded adjacent blocks that do not contain key names. 
    For instance,
        (``Slept'', `` '', 
        ``\{\{duration\}\}'', ``.'', `` '', ``Sleep'', `` '', ``early'')
        becomes three blocks
        (``Slept XXXX,
        ``\{\{duration\}\}'', ``.XXXX XXXXSleep XXXXearly''),
        and
        (``6'', `` '', ``hrs'') becomes a single block (``6XXXX XXXXhrsXX''), 
            where 
                ``X'' denotes padding.

\emparagraph{Security analysis.}
The protocol described above meets the passive security definition 
    in \S\ref{s:security-definitions};
Appendix~\ref{a:passive-security-proof} contains a proof. 
    But, at a high-level, 
        the protocol encrypts OAuth tokens and trigInp with the trigger
            service's public key so that
        they remain hidden from both \sys servers. 
    The protocol also only reveals, to each \sys server, 
        secret-shares of large blocks that are formed by 
            padding and concatenating smaller blocks;
        this leakage is captured in the 
        definition in \S\ref{s:security-definitions}.

Notably, unlike prior work~\cite{fernandes2018decentralized,
fernandes2017decoupled}, \sys's protocol ensures 
        that the trigger and action service
    learn strictly the same information as they do in
    IFTTT.
    In particular, the trigger service handles request to the trigger API; 
        it learns neither about the action API
    nor how trigOut is used to invoke the action API.
    Similarly, the action service learns only the actInp parameter needed
    to execute the action API. 
    Specifically, padding both trigOut and actInp in the same way ensures that the 
        action service cannot 
    outrightly distinguish parts
        of a string that come from trigOut versus setup-time value
            of actInp (just how it is in IFTTT).
    Thus, \sys does not increase data liability at the services---in line 
        with the data minimization or least privilege principle~\cite{gdpr,
ccpa, saltzer1975protection,
denning1976fault}.

\emparagraph{Performance analysis.}
The setup phase runs once and thus its cost gets amortized across multiple
executions of an applet.
    Here, we focus on the costs of the other phases.

In the trigger-polling phase, 
    the use of encryption and secret-sharing adds
    \cpu and network overhead at the platform
    and the services. However, this overhead is small, 
    as encryption and secret-sharing are cheap operations which
    take less than a millisecond (\S\ref{s:eval:microbenchmarks}).

In the action-generation phase, if the protocol
    invokes Yao's protocol, then the \cpu and
    network overhead for \sys's two servers increases significantly, as
    Yao represents the filterCode as a 
    low-level, verbose Boolean circuit (a network of Boolean
gates such as AND and XOR), and incurs cryptographic operations for each
gate in the circuit~\cite{lindell2009proof, zahur2015two, songhori2015tinygarble}. 
    However, 
        majority of the time, \sys calls the \texttt{string\_sub} protocol, 
        which performs cheap local computation. Indeed, in the top-50 popular applets in
                the dataset of Mi et al.~\cite{mi2017empirical},
        only one uses custom filterCode while the others use string
substitution. Similarly, in the 15 randomly selected applets we implement
(\S\ref{s:eval:compatibility}), none require custom filterCode. 

The action-execution phase is also efficient like the trigger-polling phase as
    it uses the cheap decryption and secret-sharing operations.

\section{Design for active security}
\label{s:designactive}
\label{s:design-active}

\begin{figure*}[t]
\hrule
\medskip
{ 

\begin{center}
\textbf{\Sys's protocol for the active security setting}
\vspace{-2mm}
\end{center}

\begin{myitemize5}

\item This protocol assumes each \sys server $S_b$
    has a general-purpose machine $M_b$ and three TEEs
            $T_b^{(0)}$,
        $T_b^{(1)}$, and $T_b^{(2)}$.

\item The protocol assumes that each entity has a key-pair that it uses to sign
        messages and allow others to verify signed messages.

\end{myitemize5}

\begin{center}
\vspace{-2mm}
\textbf{\textit{Setup phase}}
\vspace{-2mm}
\end{center}

        \begin{myenumerate4}
        \item The user runs steps~\ref{e:encryptoauthtokens}
                to~\ref{e:secretshareafCT} in
            Figure~\ref{fig:forttt-passive-protocol} to create $\textrm{App}_b = \{T,
                A_b, fc\}$, for all $b \in \{0, 1\}$.

        \item (Sign request to trigger API) 
        \label{e:signtrigger}
        The user appends a unique
        identifier $TID$ to $T$, and asks $TS$ to sign 
            $T$. The user obtains
        $\sigma_{T} \leftarrow \sign(sk_{TS}, T)$.

        \item 
            The user adds $\sigma_{T}$ and public key of trigger service $pk_{TS}$ to
        \label{e:sendapplettotees}
$\textrm{App}_b$,
and sends $\textrm{App}_b$ to each of $M_b$, 
            $T_b^{(0)}$,
        $T_b^{(1)}$, and $T_b^{(2)}$.
        
        \end{myenumerate4}

\begin{center}
\vspace{-2mm}
\textbf{\textit{Trigger-polling phase}}
\vspace{-2mm}
\end{center}

        \begin{myenumerate4}
        \setcounter{enumi}{3}

        \item (Call trigger API) 
            \label{e:calltriggerapi}
            $M_0$ sends $(\textrm{App}_b.T, \textrm{App}_b.\sigma_{T})$ to trigger service $TS$.
    
        \item (Generate signed shares of trigOut) 
        \label{e:generatesignedtrigout}
            $TS$ verifies $\sigma_{T}$ and
            runs step~\ref{e:generatetriggeringredients} in 
                Figure~\ref{fig:forttt-passive-protocol} to get
                $sh_b^{(trigOut)}$. 
        $TS$ then samples a unique identifier $RID$ for response, constructs
            $tout_b \gets (RID, TID, sh_b^{(trigOut)})$, and 
            computes $\sigma_{tout_b} \leftarrow \sign(sk_{TS}, tout_b)$
            $\forall b \in \{0, 1\}$.

        \item 
        \label{e:sendsignedtrigout}
        $TS$ sends $(tout_b, \sigma_{tout_b})$
            to $M_b$. 
        
        \end{myenumerate4}

\begin{center}
\vspace{-3mm}
\textbf{\textit{Action-generation phase}}
\vspace{-2.5mm}
\end{center}

        \begin{myenumerate4}
        \setcounter{enumi}{6}

            \item (Generate signed actInp shares) $M_b$ 
                sends 
                $(tout_b, \sigma_{tout_b})$ to 
                each TEE $T_b^{(i)}$, who does the following:

                \begin{myenumerate3}

                    \item 
                    \label{e:matchTID}
                        Checks $tout_b$ is from the correct $TS$ and for the
                        correct trigger API:
                
                        \begin{myenumerate4}

                            \item Asserts $\verify(\textrm{App}_b.pk_{TS}, tout_b,
                                \sigma_{tout_b})$ is True.

                            \item Asserts 
                                $TID$ in $tout_b$ equals
                                $\textrm{App}_b.T.TID$.

                        \end{myenumerate4}

                  \item (Generate actionFields shares) 
                    \label{e:generateaftees}
                    Runs step~\ref{e:generateactionfields}
                         in Figure \ref{fig:forttt-passive-protocol} with
                            $T_{1-b}^{(i)}$ to generate $sh_b^{(actInp_{i})}$. 

                  \item (Sign request to action service) 
                    \label{e:signrequesttoactionservice}
                    Constructs $ain_b^{(i)} \gets
                (tout_{b}.RID, \textrm{actionEndpoint}, C^{at \mhyphen AS},
sh_b^{(actInp_{i})})$, where
\textrm{actionEndpoint} and $C^{at \mhyphen AS}$  are part of 
$A_b$ in $\textrm{App}_b$. 
                        Then, $T_b^{(i)}$ computes 
                    $\textrm{proof}\mhyphen {T_b^{(i)}} \leftarrow
\sign(sk_{T_b^{(i)}}, ain_b^{(i)})$, gives 
                    ($\textrm{proof}\mhyphen {T_b^{(i)}}, ain_b^{(i)})$ 
        to $M_b$.

                \end{myenumerate3}

        \end{myenumerate4}

\begin{center}
\vspace{-2mm}
\textbf{\textit{Action-execution phase}}
\vspace{-2.5mm}
\end{center}

    \begin{myenumerate4}
        \setcounter{enumi}{7}
        \item 
        \label{e:callactionapi}
        (Send request) $M_b$ sends
        $ain_b^{(i)}$
from any one of its TEEs to $AS$. 
            $M_b$ also sends
            $\textrm{proof}\mhyphen{T_b^{(i)}}$ for all $i \in \{0, 1, 2\}$. 

        \item (Check proofs) 
        \label{e:checkmatch}
        For all $b \in \{0, 1\}$, and for all $i \in \{0, 1, 2\}$,
        $AS$ asserts $\verify(pk_{T_b^{(i)}},
            ain_b^{(i)},
\textrm{proof}\mhyphen{T_b^{(i)}})$ equals True.

        \item 
        \label{e:executeaction}
        $AS$ checks $RID$ in $ain_0^{(i)}$ and $ain_1^{(i)}$ is equal and unique, 
        reconstructs actInp,
validates OAuth token,
and executes action API.

    \end{myenumerate4}
}
\hrule
\caption{%
\Sys's protocol for the active security setting. This protocol uses the TEE
machines at \sys's two servers.
}
\label{fig:forttt-active-protocol}
\label{f:forttt-active-protocol}
\end{figure*}

\Sys's protocol in the previous section
    assumes a passive (honest-but-curious) adversary who can 
    monitor protocol messages but not tamper with protocol execution.
In particular, the protocol in Figure~\ref{f:forttt-passive-protocol} assumes:

    \begin{myenumerate4}[(i)]

        \item machine $M_0$ does not corrupt 
        the trigger part of the applet $T$ before sending
        it to trigger service (step~\ref{e:calltriggerfunction} in
Figure~\ref{f:forttt-passive-protocol}), 

        \item 
            $M_0$ and $M_1$ do not tamper 
            with trigOut or the applet parts that
            feed into action-generation (step~\ref{e:generateactionfields} in
Figure~\ref{f:forttt-passive-protocol}),

        \item 
            $M_0$ and $M_1$ run GenerateAI correctly,

        \item 
            $M_0$ and $M_1$ do not tamper with the message
        sent to the action service (step~\ref{e:executeaction} in
Figure~\ref{f:forttt-passive-protocol}), and

        \item machines $M_0$ and $M_1$ do not 
            replay requests to action service (requests to
            a trigger service 
            are idempotent).
        \end{myenumerate4}

One way to relax these assumptions is to 
    extend the protocol with zero-knowledge proofs~\cite{setty2018proving,
walfish15verifying, parno13pinocchio} that allow
    proving the correctness of code execution without revealing
    the sensitive data that the code is operating on.
    However, these protocols are expensive. 
    Instead, \sys uses
        digital signatures and trusted execution environments (TEEs)---recall, in the active setting,
        each \sys server has three TEE machines (\S\ref{s:arch}).
At a high-level, 
signatures enable services to verify that
    that the information they are receiving has not been tampered. 
Meanwhile, TEEs such as Intel SGX~\cite{sgx}, AMD SEV~\cite{kaplan2016amd}, ARM
TrustZone~\cite{trustzone}, 
    provide a tamper-proof sandbox for executing computation. 
Figure~\ref{fig:forttt-active-protocol} shows \sys's protocol that incorporates
    these integrity primitives;
we now briefly describe each phase and reflect on the protocol's security
guarantees and
performance.

\emparagraph{Setup and trigger-polling.}
The main difference in these phases (relative to the passive security protocol)
    is the addition of signatures
(steps~\ref{e:signtrigger} and~\ref{e:generatesignedtrigout} in
Figure~\ref{f:forttt-active-protocol}) 
    to the input and output of the trigger API.
Besides signatures, these phases add
    unique IDs, particularly to the output of the trigger API. 

\emparagraph{Action-generation.}
This phase uses the TEE machines at the two servers.
    Each TEE (i) matches the response from the trigger service 
    to the correct applet (step~\ref{e:matchTID} in
Figure~\ref{f:forttt-active-protocol}), (ii) runs
    the GenerateAI function, and (iii)
    signs its part of the 
        request to the action service (step~\ref{e:signrequesttoactionservice}
in Figure~\ref{f:forttt-active-protocol}).
    This signature becomes a proof
        that the action service should act on the request.
    However, since all but one of the TEE machines may be compromised
        (\S\ref{s:security-definitions}), the action service
        acts on the request only if proofs from all the TEEs are valid.

\emparagraph{Action-execution.}
The action service checks the proofs, 
    ensures 
        the uniqueness of the unique identifier in the request,
    and executes the action API.

\emparagraph{Security analysis.}
The protocol described in Figure~\ref{f:forttt-active-protocol}
    meets the active security definition in \S\ref{s:security-definitions}.
    Appendix~\ref{a:active-security-proof} contains a proof. 
    But, briefly, the cryptographic binding guarantees of digital signatures ensures 
        that the messages the services receive are not tampered.
    Besides, the use of TEEs and equality checks ensures that 
        actInp are computed correctly.
    Finally, the use of unique IDs ensures that requests
        to action service are not replayed.
    
\emparagraph{Performance analysis.}
As with the protocol for passive security, the cost of setup phase
    gets amortized across many invocations of an applet. Therefore, 
    we focus on the costs of the other phases.

The protocol adds (relative to the passive security protocol) low \cpu and network overhead
    in the trigger-polling phase, for two reasons.
    First, signature generation and checking takes less than a millisecond
(\S\ref{s:eval:microbenchmarks}), and each signature
    is 64 bytes with elliptic-curve cryptography. 
    Second, the trigger service sends its output (step~\ref{e:sendsignedtrigout}
        in Figure~\ref{f:forttt-active-protocol})
        once to each server; in contrast, a naive 
            design would make the trigger service send a copy
            of trigOut to each TEE separately.

In the action-generation phase, 
    the main overhead is due to 
    the replication of GenerateAI function three times
    on the three TEEs (step~\ref{e:generateaftees} in
Figure~\ref{f:forttt-active-protocol}). When
    GenerateAI further calls Yao's protocol, this
    increase is significant as Yao's \cpu and network costs are high.
    However, when GenerateAI calls \sys's efficient
        \texttt{string\_sub} procedure, then the impact on the protocol is
            small.

In the action-execution phase, the protocol adds
        low \cpu and network overhead.
        Again, the protocol sends only one copy of actInp
        from each server to the action service. 
        Each TEE does send a separate signature but 
        signatures are small in size (64 bytes each) and signature
         verification takes less than a millisecond of \cpu time.

\section{Supporting OAuth tokens with expiry}
\label{s:refreshtokens}
\label{s:refresh-tokens}

So far, we have assumed that OAuth tokens do not expire (the
user puts tokens in applet during the setup phase and goes offline).
    However, in practice, some IFTTT applets use OAuth
    that do expire. 
This section first
quickly reviews how 
        expirable tokens work, and 
then describe \sys's protocol.

\emparagraph{Expirable tokens in OAuth protocol.}
The OAuth protocol uses two tokens: an access token and a refresh
token. Each token has an expiration time. 
    The holder of these tokens includes the access token in regular GET and POST requests to
    HTTP endpoints. 
    However, if the access token is expired, the endpoint returns a HTTP 401
    unauthorized error. 
    Then, the token holder sends a \emph{refresh} request to an
    OAuth endpoint
    consisting of both the access token and the refresh token; if
    the refresh token has not expired, the endpoint returns a new
    access token and refresh token~\cite{hardt2012oauth}.

\emparagraph{\Sys's protocol to support OAuth token refresh.}
One way for \sys to support expirable tokens is to 
    use the TEEs at the two servers, who could initiate the refresh tokens
    request and update the 
        tokens in an applet. 
However, \sys uses a protocol that avoids the use of TEEs. 
\Sys's high-level idea is to use a signed
    chain of encrypted tokens---to link the
    original user-supplied tokens with new tokens.

Assume that time is split into epochs, beginning with the $0$-th epoch.
Also, let $\textrm{at}_k$ be the valid access token in the $k$-th epoch, and
            define $(\enc(pk, \textrm{at}_0 \| \textrm{at}_k), \sigma_k)$
    as the \emph{token-chain} for the same epoch; here,
        encryption is under the key of the entity issuing the tokens (a
        trigger or action service) and $\sigma_k$ is a signature by the same
        entity over the ciphertext. 
Then, during a regular HTTP request to a trigger or action API in
the $k$-th epoch 
(steps~\ref{e:calltriggerapi} and~\ref{e:callactionapi} in
Figure~\ref{f:forttt-active-protocol}),
machines
$M_0$ and $M_1$ at \sys's platform append the $k$-th token-chain to the request.
When the service receives the request, it verifies the signature, and
        matches the $\textrm{at}_0$ part of the chain
        to the original user-supplied access token in the request.
    If there is a match, then the service replaces the original token in the
        request with $\textrm{at}_k$ (this replacement is done in the
            API-agnostic logic added to the services; \S\ref{s:arch}).

A question is: how does \sys's machines $M_0$ and $M_1$ 
     obtain the token-chain for the
    $k$-th epoch?
The user supplies the chain for the $0$-th epoch, that is,
            $(\enc(pk, \textrm{at}_0 \| \textrm{at}_0), \sigma_0)$,
        as part of the applet during setup.
To enable $M_0$ and $M_1$ to get the chain for the $k$-th epoch, \sys extends the 
    refresh token request to the OAuth endpoint.
As mentioned above in the background, normally the token
holder sends the
        refresh and access tokens for the $k$-th epoch and receives
        refresh and access tokens for the $(k+1)$-th epoch. 
In \sys, $M_0$ includes the token-chain for the the $k$-th epoch, that is, 
            $(\enc(pk, \textrm{at}_k \| \textrm{at}_0), \sigma_k)$,
    in the refresh request. 
The service 
    verifies that the
            token $\textrm{at}_k$ in the chain matches the
        token in the regular part of the refresh request, and sends
    the chain for the $(k+1)$-th epoch, 
    that is,
            $(\enc(pk, \textrm{at}_{k+1} \| \textrm{at}_0), \sigma_{k+1})$.

\emparagraph{Security analysis.}
\Sys's protocol with the extension described above meets the 
    active security definition in \S\ref{s:security-definitions}.
Appendix~\ref{a:active-security-proof} contains a proof.
    But, briefly, 
    \sys's protocol keeps tokens (both original and in the chain) for all epochs encrypted with the services' keys.
    Also, the \sys platform makes
        a successful request to a trigger and action API if it has
            a valid chain and a valid (but expired) user-supplied token.
    Finally, \sys's servers get a chain for a new epoch only
        if they possess the expiring (but valid) chain.

\emparagraph{Performance analysis.}
The protocol extension affects mainly the trigger-polling and action-execution
phases. 
    In particular, the addition of token-chain to the requests to the
    trigger and action APIs 
(steps~\ref{e:calltriggerapi} and~\ref{e:callactionapi} in
Figure~\ref{f:forttt-active-protocol})
        adds
    \cpu at the services to verify the chain, and network
        transfers between the platform and services to transport the chain.
Meanwhile, the refresh tokens request (to get a chain for a new epoch) adds
    token chains but is made infrequently.

\section{Implementation}
\label{s:impl}
We have implemented a prototype of \sys (\S\ref{s:arch},
\S\ref{s:design-passive}--\S\ref{s:refresh-tokens}).
Our prototype is written in Python and C. 
The Python part implements 
    a web server that interacts with the user and the services.
The C part implements
    RPC servers inside TEEs to execute
        the action-generation step of \sys's protocol.
Both parts build on existing libraries and frameworks, including        
        Flask~\cite{flask} for the web server,
            MongoDB~\cite{mongodb} for
        storing applets,
            Authlib~\cite{authlib} for the OAuth protocol, 
    OblivC~\cite{zahur2015obliv} for Yao's garbled circuits,
    OpenSSL~\cite{openssl} for cryptographic functions,
        and SGX-LKL~\cite{sgx-lkl} as a library operating system.
\Sys's implementation over existing libraries is
    2128 lines of Python and
2721 lines of C.\footnote{\Sys's code will be made available on GitHub
    after its publication.}

\section{Evaluation}
\label{s:eval}
\label{s:evaluation}

\Sys's evaluation answers the following questions:
\begin{myenumerate}
    \item What are \sys's overheads relative to a non-secure baseline that
            mimics IFTTT?

    \item How much overhead do \sys techniques
            (\S\ref{s:design-passive}---\S\ref{s:refreshtokens}) incur?
        
    \item For what kinds of applets is \sys practical?    
    
\end{myenumerate}

\newcommand{\hackbullet}[1]{\parbox{1.9in}{\begin{myitemize4} \item #1 \end{myitemize4}}}
\begin{figure}[t]
\footnotesize
\centering

\begin{tabular}{@{}
>{\raggedright\arraybackslash}m{2.9in}
>{\raggedright\arraybackslash}b{0.25in}
@{} }
\toprule

    For applets with custom filterCode ($\approx$1-2\%), \sys increases
platform-side \cpu and network costs
        by 54$\times$ and 1883$\times$  over a non-secure baseline.
    However, for common applets, \sys increases these costs 
    by 3.6$\times$ and
       4.3$\times$. 
    
            & \S\ref{s:eval:serversideoverheads} \\
[0.75 cm]

For all applets, \sys increases \cpu and network costs at trigger and action services 
    by up to 3.2$\times$ and 2.52$\times$, respectively, relative to a non-secure baseline.
        & \S\ref{s:eval:servicesideoverheads} \\ [0.55cm]

\Sys's dollar cost for running the common applets is
    3.74$\times$ of a non-secure baseline. & 
    
        \S\ref{s:eval:dollarcost} \\ [0.45 cm]

One can program existing IFTTT applets on \sys with ease. &
    \S\ref{s:eval:compatibility} \\

\bottomrule
\end{tabular}

\caption{Summary of main evaluation results. 
}
\label{f:evalsummary}
\label{fig:evalsummary}
\end{figure}

\begin{figure}[t]
\footnotesize
\centering

\begin{tabular}{
@{}
*{1}{>{\raggedright\arraybackslash}b{.07\textwidth}}  @{ } %
*{1}{>{\raggedleft\arraybackslash}b{.09\textwidth}}  @{ } %
*{1}{>{\raggedleft\arraybackslash}b{.06\textwidth}}  @{ }   %
*{1}{>{\raggedleft\arraybackslash}b{.065\textwidth}} @{ }  %
*{1}{>{\raggedleft\arraybackslash}b{.05\textwidth}}  @{ } %
*{1}{>{\raggedleft\arraybackslash}b{.115\textwidth}}  @{ } %
@{}
}

machine    & vendor  & CPUs  & RAM   & SGX  & location \\
\midrule
B1ms        & MS Azure       &   1   & 2~GB     & No  & SF, CA \\
DC4s\_v2    & MS Azure       &   4   & 16~GB    &   Yes & Blue Ridge, VA \\

B1.1x2      & IBM Cloud      &   1   & 2~GB     & No  & SJ, CA \\
bm          & IBM Cloud      &   4   & 16~GB    &   Yes & SJ, CA  \\
n1-std-4 & Google Cloud &   4   & 15~GB    & No  & The Dalles, OR \\
\bottomrule
\end{tabular}
\caption{%
Machines used in \sys's experiments.
}
\label{fig:testbed}
\label{f:testbed}
\end{figure}

\noindent Figure~\ref{f:evalsummary} summarizes our evaluation results. 

\emparagraph{Method and setup.}
We compare \sys to the following baseline systems.
\emph{NoSec} is our Python-based reference implementation of
        IFTTT 
  (IFTTT itself is closed-source). NoSec uses Flask as a web
        server, MongoDB as a data store for applets, and the
        Python requests framework~\cite{requests} for HTTP requests.
\emph{W-Yao}, \emph{W-C}, and \emph{W-I} are intermediate
    baselines between NoSec and \sys.
    W-Yao implements \sys's protocol
        for the passive security setting. It provides
        confidentiality but no integrity guarantees.
        Further, it uses Yao's protocol only
        for secure computation (\S\ref{s:design-passive}).
    W-C builds on W-Yao and adds \sys's custom protocol
        for string substitution for the common types of applets on IFTTT 
        (\S\ref{s:design-passive}).
    W-I builds on W-C and adds the use of digital signatures and TEEs
        for integrity guarantees (\S\ref{s:design-active}).
    \Sys (W) builds on W-I and adds the use of token-chains for 
        refreshing OAuth tokens (\S\ref{s:refreshtokens}).

We perform a series of experiments to answer the evaluation questions.
In each experiment, we deploy a system, execute an applet (described below),
    and measure
        \cpu time, wall-clock time, network transfers, and storage space use.
    We measure \cpu time and network transfers using
        kernel accounting frameworks \texttt{/proc/pid/stat} and
        \texttt{/proc/net/dev}, respectively.
    We measure wall-clock time by instrumenting code with Python's
        \texttt{time.time()}, and storage space use by reading
        statistics returned by MongoDB's \texttt{collStats}. 

We execute the ``Get a daily 6:00 AM email with the weather report'' applet
    from Weather Underground~\cite{wuemail}.
  We choose this applet because it is both popular (installed over 30K times) and
        representative of the common IFTTT applet.
    We run this applet with three types of filterCode.
        The ``passing around'' filterCode sets the body of the email
            to the value of the key-value pair with key ``new\_weather\_type''
            in trigOut.
    The ``string sub'' filterCode sets the body of the email by
        substituting the placeholder
        in the template string
        ``This is an example of a substituted string. The new type of weather is \{\{new\_weather\_type\}\}'' with the
            key-value pair with key ``new\_weather\_type'' in trigOut. 
                    The ``custom code'' filterCode
            sets the email body to a variant of the string above depending on the type of
                weather (``sunny'' versus ``rainy''). 
    For \sys and its variants, we pad the template strings
        and the values in trigOut as described in
        Section~\ref{s:design-passive}.
    We configure \sys to pad each block in a string to a power-of-two length.

Our testbed (Figure~\ref{f:testbed}) is a set of VMs on Google Cloud, Microsoft Azure, and
IBM Cloud. 
    We run the trigger and action services on Google Cloud, and \sys's two
        servers on Azure and IBM. 
    Within Azure and IBM, we use both general-purpose and TEE machines.
        Cloud providers currently
            offer only Intel SGX-based TEEs~\cite{sgxavailability}; we use three 
            such machines
            as the three TEEs for W-I and \sys.
    However, the TEE availability on cloud providers is improving~\cite{sgxavailability, awsnitro}, and
         in a real deployment, \sys would use three different TEEs.
        
\begin{figure}[t]
    \footnotesize
    \centering
    
    \begin{tabular}{
        @{}
        *{1}{>{\raggedright\arraybackslash}b{.07\textwidth}}  @{}%
        *{1}{>{\raggedleft\arraybackslash}b{.07\textwidth}}  %
        *{1}{>{\raggedleft\arraybackslash}b{.07\textwidth}}  %
        *{1}{>{\raggedleft\arraybackslash}b{.07\textwidth}}  %
        @{}
        }
        & \cpu time & \cpu time & \cpu time   \\
        & for 10B & for 100B & for 1000B             \\
        \midrule

        \multicolumn{1}{@{}l}{\textbf{secret sharing-related}} \\

        \multicolumn{1}{@{}l}{XOR share}       &  5.7 $\mu$s &  21.0 $\mu$s  &  178.0 $\mu$s \\
        \multicolumn{1}{@{}l}{XOR reconstruct} &  5.2 $\mu$s &  32.0 $\mu$s &  306.0 $\mu$s \\

        \midrule 

        \multicolumn{1}{@{}l}{\textbf{encryption-related}} \\

        \multicolumn{1}{@{}l}{ECIES encrypt} &  473.0 $\mu$s &  491.0 $\mu$s &  498.0 $\mu$s   \\
        \multicolumn{1}{@{}l}{ECIES decrypt} &  486.0 $\mu$s &  491.0 $\mu$s &  514.0 $\mu$s   \\

        \midrule

        \multicolumn{1}{@{}l}{\textbf{signature-related}} \\

        \multicolumn{1}{@{}l}{ECDSA sign} &  685.0 $\mu$s &  692.0 $\mu$s  &  695.0 $\mu$s   \\
        \multicolumn{1}{@{}l}{ECDSA verify} &  550.0 $\mu$s &  555.0 $\mu$s &  559.0 $\mu$s   \\

        \midrule
    \end{tabular}

    \begin{tabular}{
        @{}
        *{1}{>{\raggedright\arraybackslash}b{.05\textwidth}}  @{}%
        *{1}{>{\raggedleft\arraybackslash}b{.12\textwidth}}  %
        *{1}{>{\raggedleft\arraybackslash}b{.12\textwidth}}  %
        @{}
        }
        & \cpu time & network transfers \\
        
        \midrule

        \multicolumn{1}{@{}l}{\textbf{Yao-related}} \\

        \multicolumn{1}{@{}l}{$\phi$ = integer comparison} &  90.0 $\mu$s &   6.2 KB  \\

        \multicolumn{1}{@{}l}{$\phi$ = string substitution} &   294.3 ms &   862.1 KB  \\

        \bottomrule

    \end{tabular}

    \normalfont\selectfont
    \caption{%
    \cpu times and network transfers for cryptographic operations in \sys and
the baselines.
    The key-sizes of elliptic curve-based ECIES and ECDSA are 256 bits. The
elliptic curve used is secp256k1.
    The overheads of Yao depend on the function being computed inside Yao;
        integer comparison compares two 32-bit integers,
        and the string substitution function replaces the 
            braces in
        the template string ``This is a substitution \{\}'' with another string of length
100 bytes. 
    }
    \label{fig:microbenchmarks}
    \label{f:microbenchmarks}
\end{figure}

\begin{figure*}[t]
\centering
\includegraphics[width=3.47in]{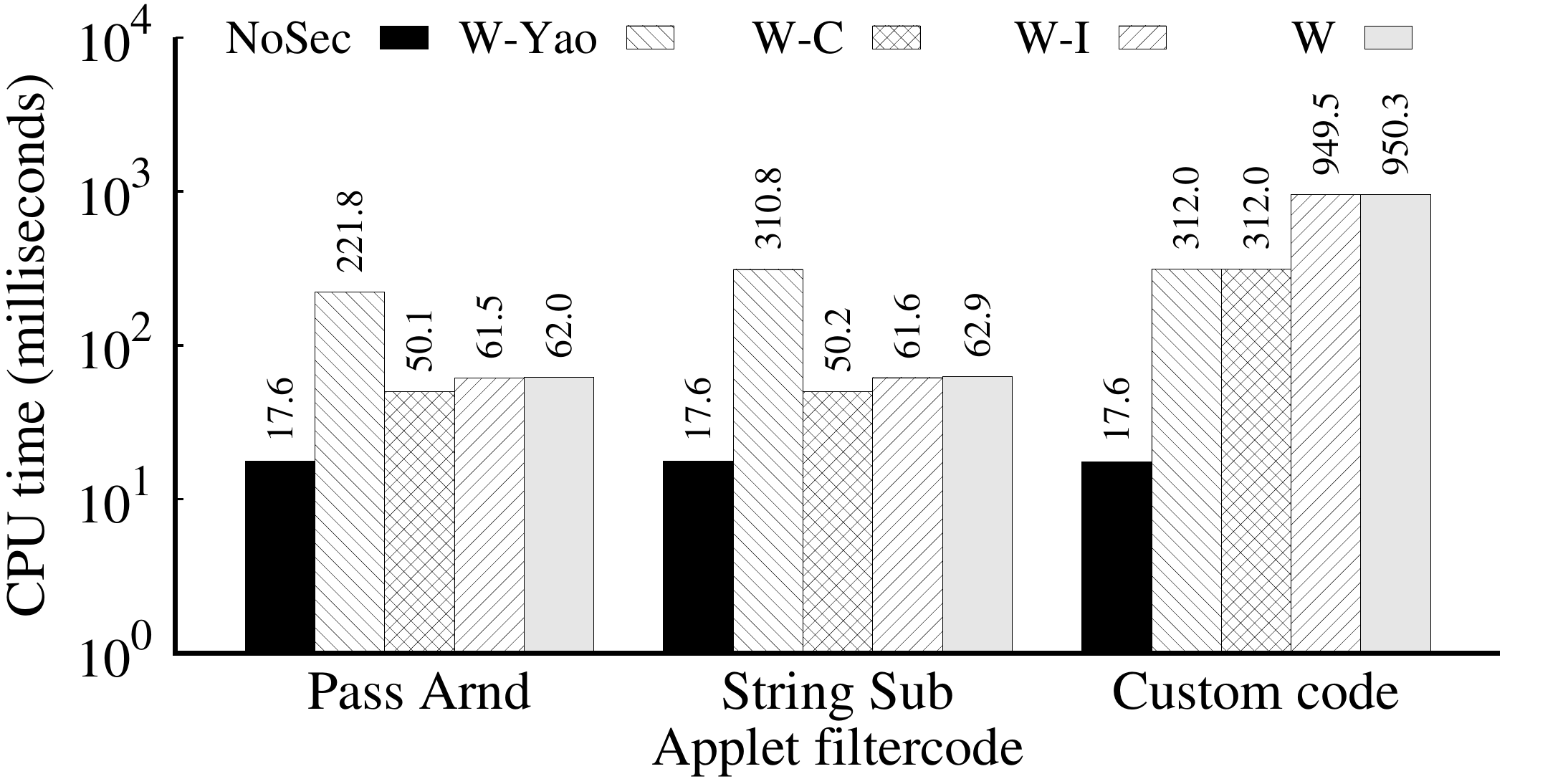}
\includegraphics[width=3.47in]{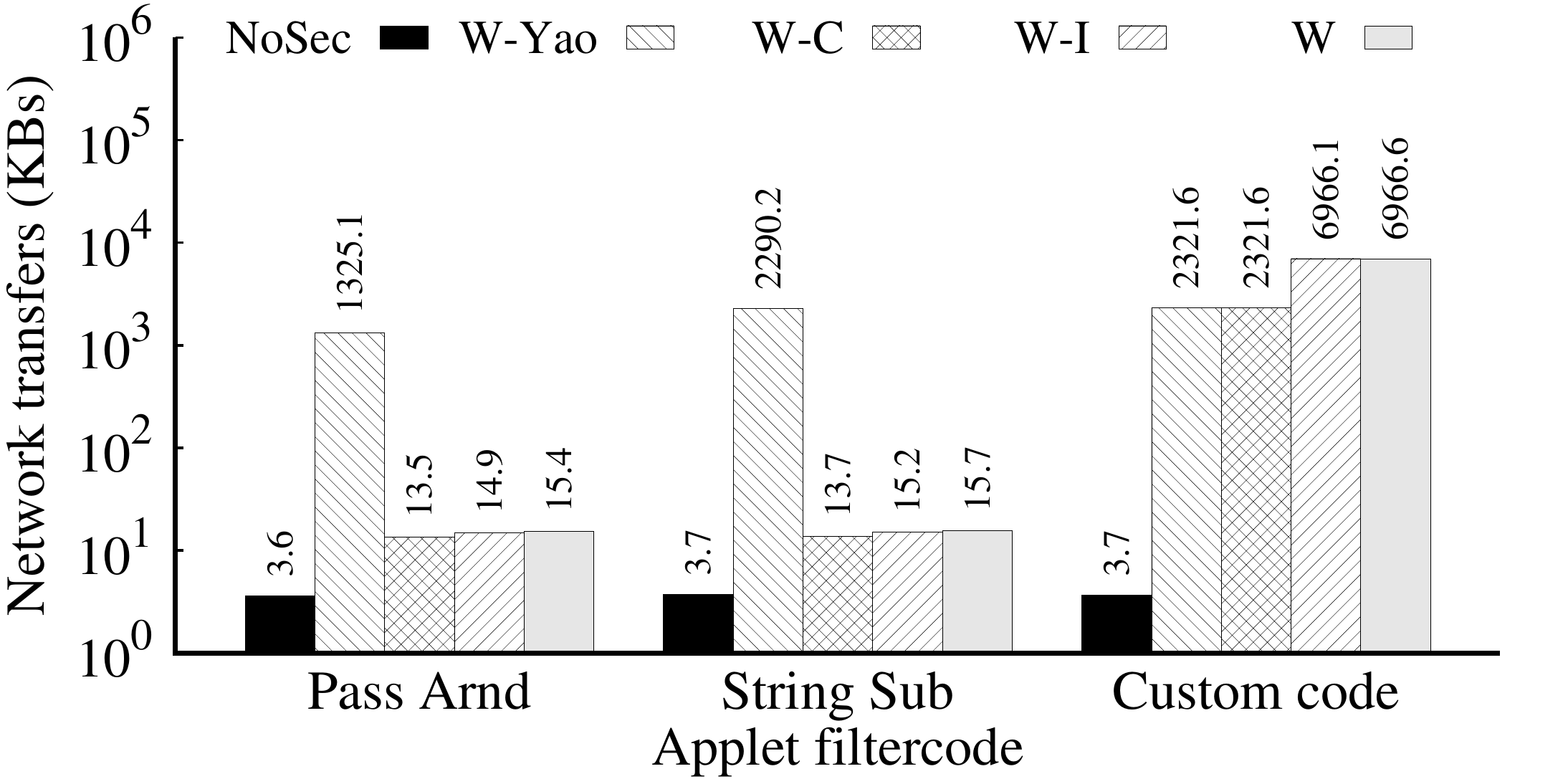}
\caption{Platform-side \cpu time and network transfers 
        for \sys and the baseline systems for different filterCodes.
}
\label{f:servercpunetwork}
\label{f:servercpu-run}
\label{f:servernetwork-run}
\end{figure*}

\emparagraph{Microbenchmarks.}
\label{s:eval:microbenchmarks}
Figure~\ref{fig:microbenchmarks} shows \cpu and network costs
    for common cryptographic operations in \sys and the baselines
    on machines of type B1.1x2 (Figure~\ref{f:testbed}).
\Sys uses elliptic curve-based encryption and signature schemes (ECIES and
ECDSA, respectively).
We will use these microbenchmarks to explain the numbers in the next
subsections.

\subsection{Platform-side overheads}
\label{s:eval:serversideoverheads}
\label{s:eval:platformsideoverheads}
\emparagraph{\cpu overhead.}
Figure~\ref{f:servercpunetwork}~(a) shows platform-side \cpu overhead while
varying the system and the filterCode in our workload applet.    
At a high level,
    the \cpu overhead increases
        as the complexity of filterCode increases.
    For instance, \sys's \cpu time is
            62.0~ms for pass arnd,
            62.9~ms for string sub, and
            950.3~ms for custom filterCode. 

For all three filterCode types,
    NoSec consumes the least \cpu, while
        W-Yao and \sys consume the most.
For instance, for the string sub filterCode,
    W-C, W-I, \sys, and W-Yao consume
    2.9$\times$, 3.5$\times$, 3.6$\times$, and 17.7$\times$
    the \cpu time of NoSec. 
This trend is expected as
    NoSec operates over plaintext data
    and does not provide integrity properties,
    while W-C adds secret-sharing, encryption,
        and tailored string substitution over shares of strings
            (Figure~\ref{f:forttt-passive-protocol} in
        \S\ref{s:design-passive}),
    W-I further adds
        signature verification
        and generation
        inside TEEs
        (Figure~\ref{f:forttt-active-protocol} in \S\ref{s:design-active}), and 
            \sys adds
            token-chain to W-I (\S\ref{s:refreshtokens}).
    W-Yao consumes high \cpu as it performs
        string substitution over an
        encrypted Boolean circuit
        (\S\ref{s:design-passive}).

Meanwhile, the \cpu times connect back to microbenchmarks
(Figure~\ref{f:microbenchmarks}).
   For instance, \sys adds 12.7~ms of \cpu time relative to W-C
        for the string sub filterCode.
    This is primarily due to three pairs of signature generation and six pairs of signature verification
        using OpenSSL inside TEE machines;
        each signature generation and verification takes $\approx$0.7~ms and
            $\approx$0.55~ms, respectively.

\begin{figure}[t]
\centering

\includegraphics[width=3.47in]{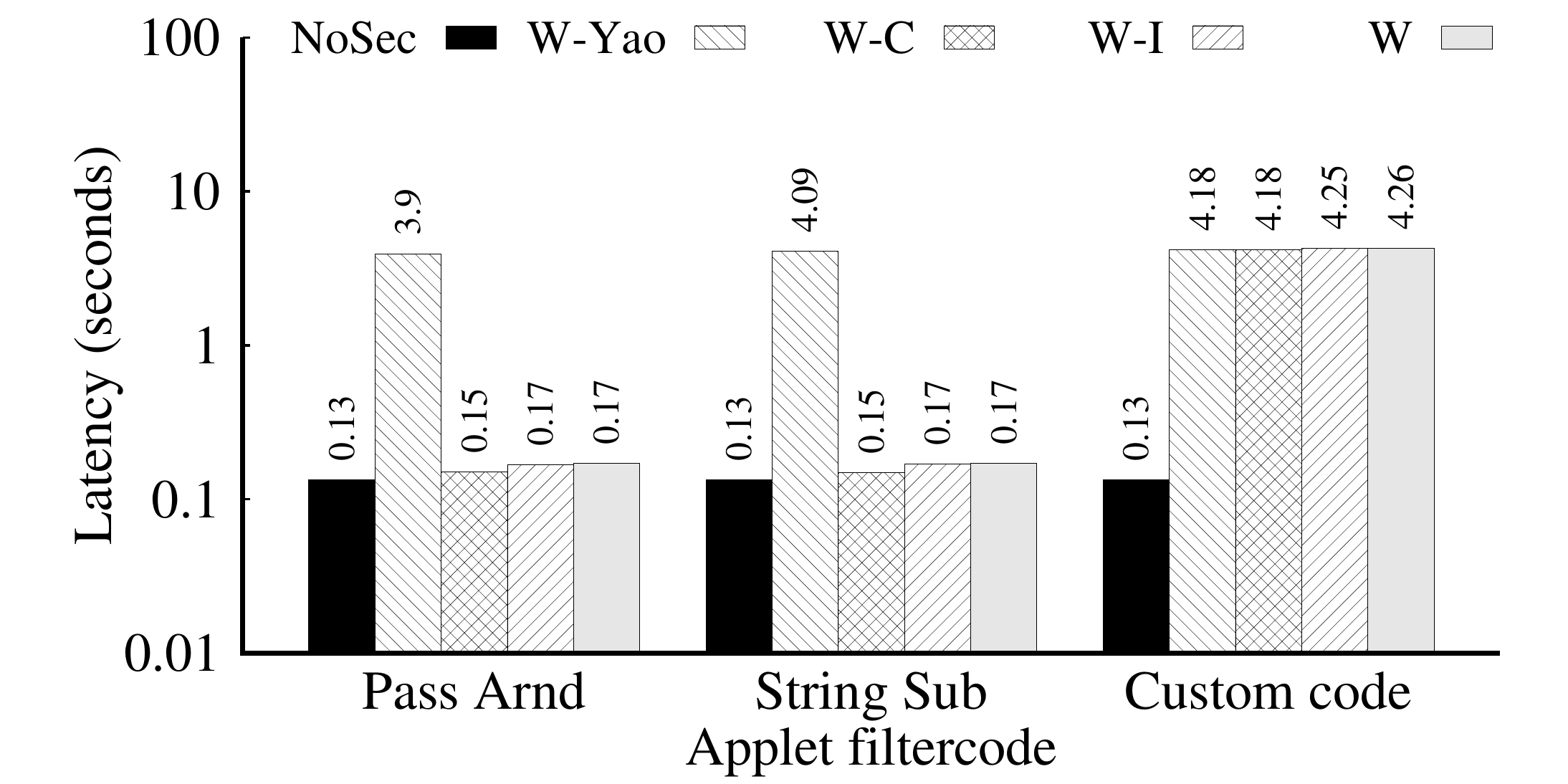}
\caption{Platform-side 
        trigger-action latency.} 
\label{f:serverstoragelatency}
\label{f:serverstorage-setup}
\label{f:serverlatency-run}
\end{figure}

\emparagraph{Network overhead.}
    Figure~\ref{f:servercpunetwork}~(b) shows platform-side network overhead
    for the various systems and filterCodes.
The network trends resemble those for \cpu.
The costs increase with filterCode complexity,
and for each type of filterCode, W-C, W-I, \sys, and W-Yao incur successively
    increasing cost over NoSec,
except for the custom code filterCode where W-C behaves like W-Yao.

\Sys incurs overhead over NoSec because it transfers
            serialized ciphertexts,
                secret-shares, signatures, and token-chains during the trigger-polling,
                    action-generation, and action-execution steps of \sys's
protocol (\S\ref{s:design-passive}--\S\ref{s:refreshtokens}).
As an example,            
    consider the trigger-polling message from \sys's first server to the trigger service. 
 In NoSec, the header of the message is 219 bytes, while it is
    383 bytes in \sys. The header increases by 164 bytes because
        the 64 byte OAuth token in NoSec is ECIES-encrypted to a 161-byte
            ciphertext,
            and then
            serialized to a final size of 228 bytes (step~\ref{e:calltriggerapi}
                    in Figure~\ref{f:forttt-active-protocol}).
Similarly, the body of the polling message is 150 bytes in NoSec and 826 bytes
    in \sys, due to the addition of several fields including
        231 bytes of encrypted trigInp,
            126 bytes of a signature over trigInp, 
        and 403 bytes of a token-chain.

\emparagraph{Latency overhead.}
Figure \ref{f:serverlatency-run}
    shows the time it takes for the various systems to execute an applet.
    Our wall-clock timing begins when the platform initiates the
trigger-polling request to the trigger
    service, and ends when the action services
        receive the action details. 
\Sys's latency is 1.3$\times$, 1.3$\times$, and 32.7$\times$ of NoSec's latency
    for the three types of filterCode. The latency increase
    for the custom filterCode is high because
        it calls the expensive Yao's protocol. 
        
\begin{figure}[t]
    \footnotesize
    \centering

    \begin{tabular}{
        @{}
            *{1}{>{\raggedright\arraybackslash}b{.08\textwidth}}  @{}%
            *{1}{>{\raggedleft\arraybackslash}b{.14\textwidth}}     %
            *{1}{>{\raggedleft\arraybackslash}b{.14\textwidth}}      %
        @{}
        }
        Component & NoSec applet & \Sys Applet  \\
        \midrule

        \multicolumn{1}{@{}l}{title, description}  &  27 bytes &   27 bytes \\
        
        \multicolumn{1}{@{}l}{filterCode  }  &  62 bytes &   139 bytes \\
        \multicolumn{1}{@{}l}{trigger     }  & 337 bytes &  793 bytes \\
        \multicolumn{1}{@{}l}{action      }  & 340 bytes &  3506 bytes \\

        \midrule
        \multicolumn{1}{@{}l}{total}  &  766 bytes & 4465 bytes \\
        \bottomrule

    \end{tabular}
    \normalfont\selectfont
    \caption{Comparison of storage space consumed by a NoSec applet
 and a \sys
applet .} 
    \label{f:eval:applet-storage-breakdown}
\end{figure}

\begin{figure*}[t]
\centering
\includegraphics[width=3.47in]{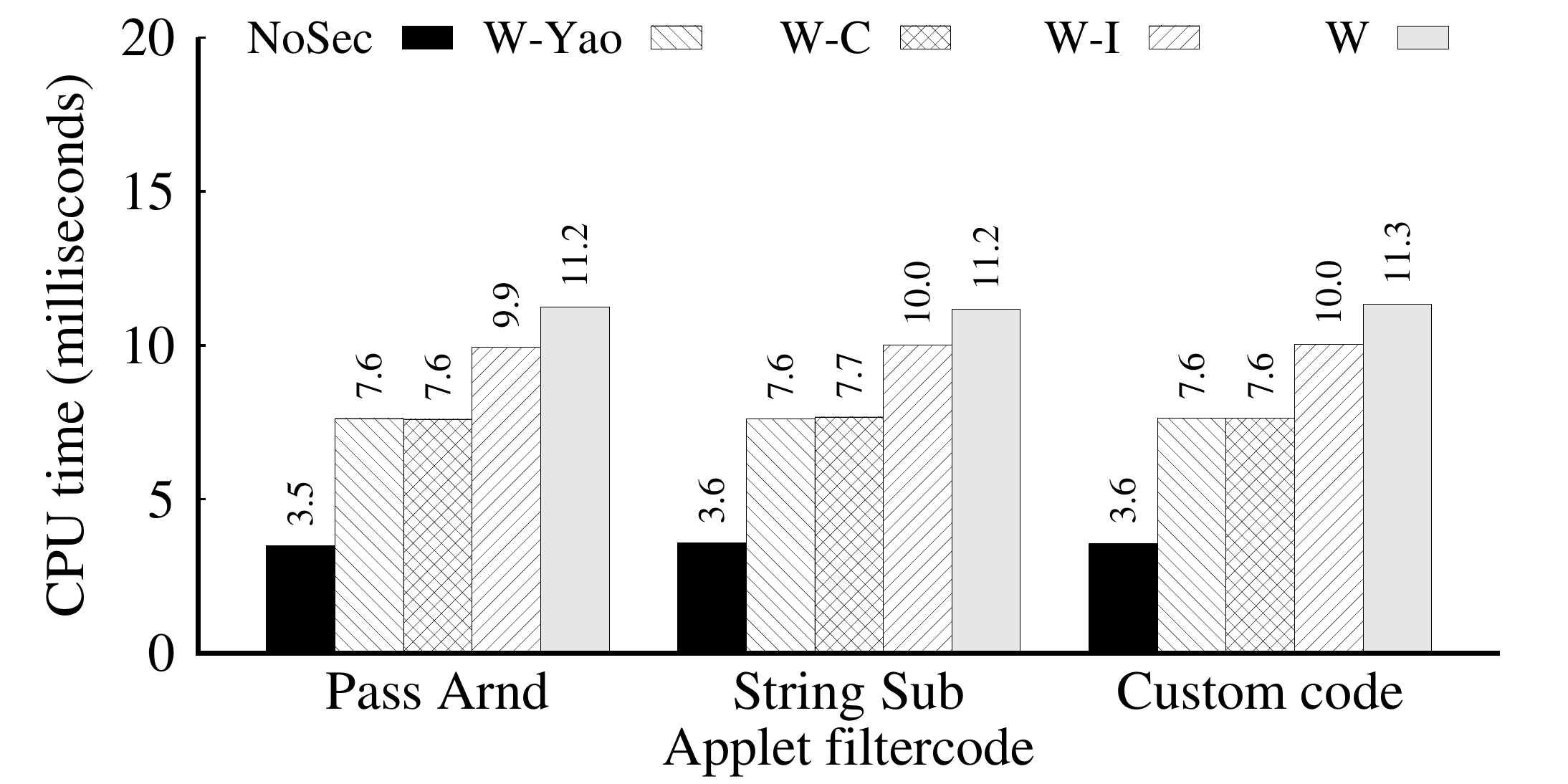}
\includegraphics[width=3.47in]{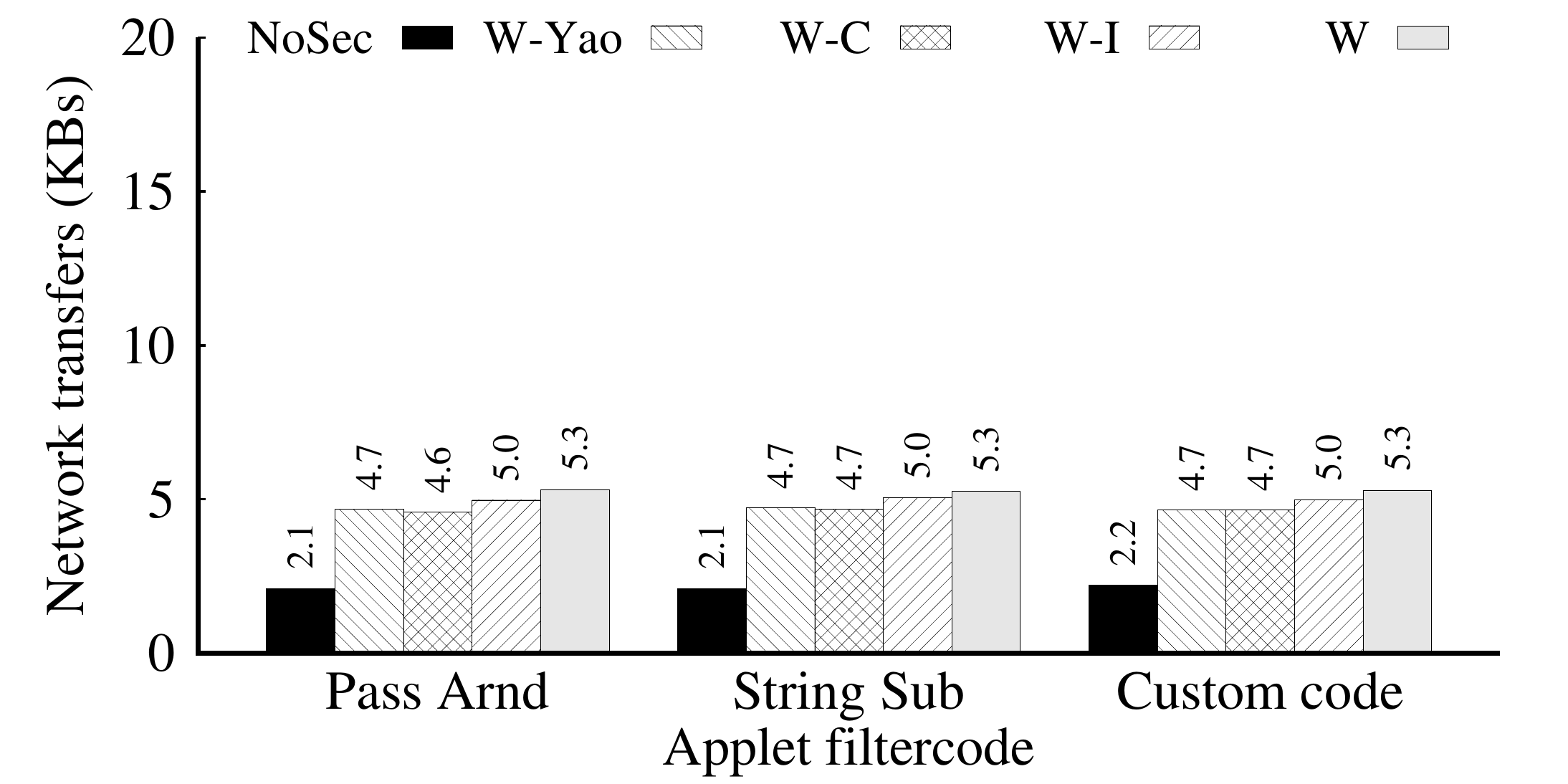}
\caption{Trigger-service \cpu time and network overhead 
        for \sys and the baseline systems for different filterCodes.}
\label{f:triggerserviceoverhead}
\end{figure*}

\begin{figure*}[t]
\centering
\includegraphics[width=3.47in]{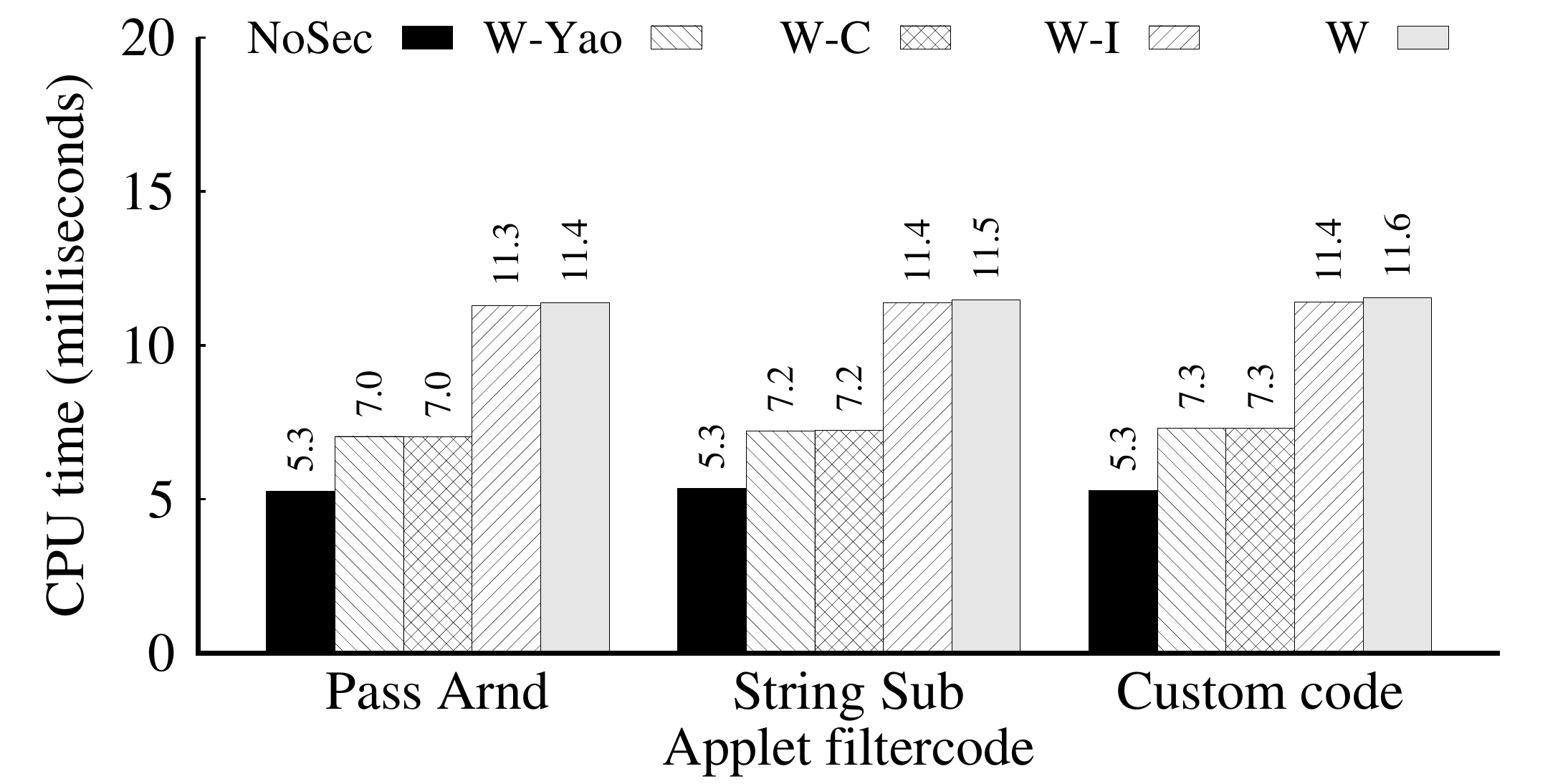}
\includegraphics[width=3.47in]{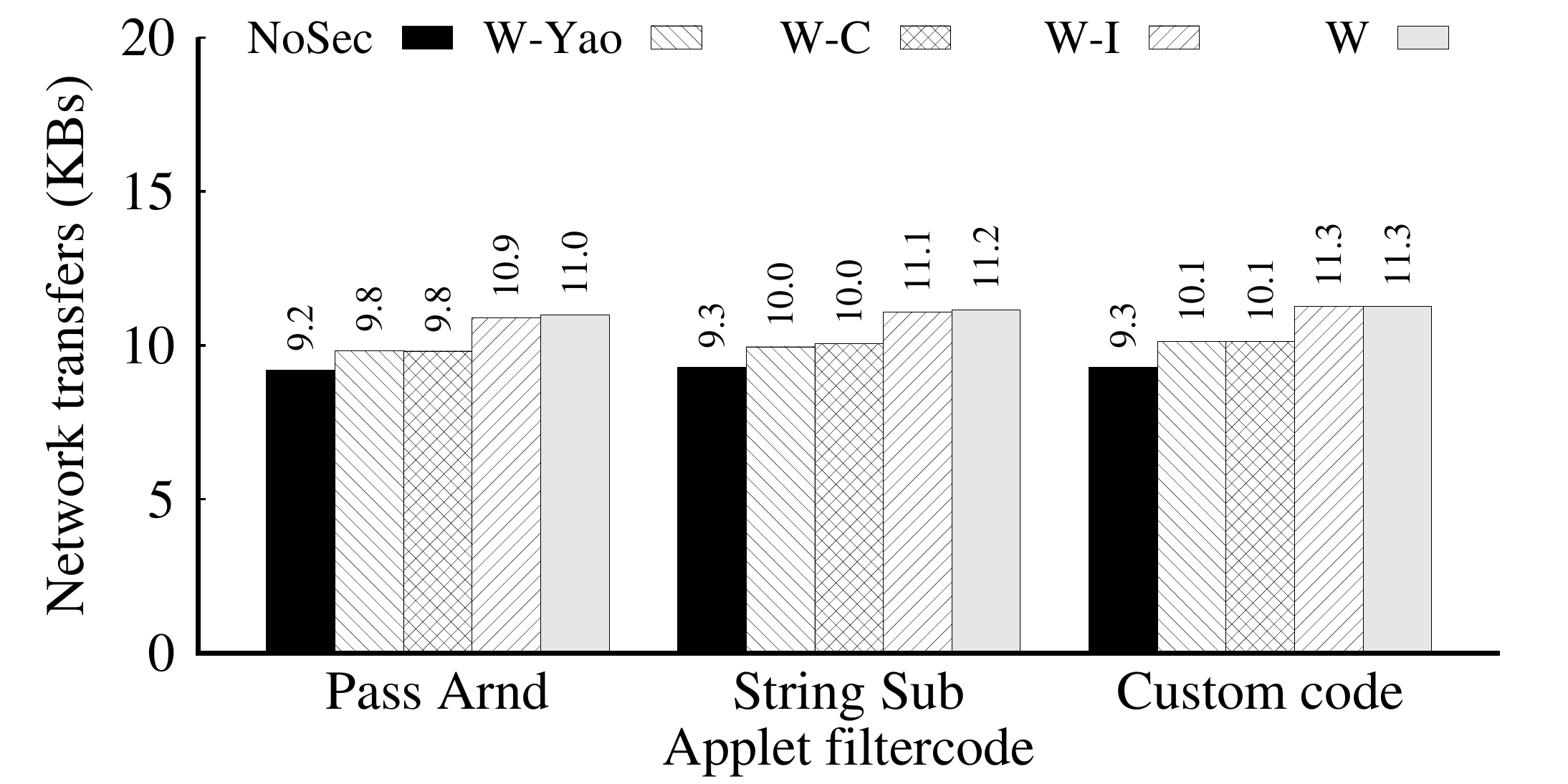}
\caption{Action-service \cpu time and network overhead 
            for \sys and the baseline systems for different filterCodes.}
\label{fig:actionservicecpu-run}
\label{f:actionservicenetwork-run}
\end{figure*}

\emparagraph{Storage overhead.}
Figure~\ref{f:eval:applet-storage-breakdown}
    shows \sys's and NoSec's storage space use
        for the applet with the string sub filterCode.
Overall, \Sys's applet takes
        5.8$\times$ the storage space required by the NoSec applet.
This extra storage is due to the additional components
    in \sys's applet, such as 
        secret-shares of setup-time values of actInp, and the overhead
        of encrypting, signing, and serializing the components.

\subsection{Service-side overheads}
\label{s:eval:servicesideoverheads}
\emparagraph{Trigger service overheads.}
Figure~\ref{f:triggerserviceoverhead}~(a)
    shows \cpu time 
        for the various systems and the three types of
            filterCodes in our workload applet.
The \cpu times have two notable aspects.
    First, \sys's \cpu time
            is $3.2\times$
    that of NoSec.
The overhead comes from
        confidentiality techniques
            (decryption of trigInp ciphertexts and
            creation of trigOut secret shares) and
        integrity design
        (verification and generation of signatures). 
Second, the overhead does not vary with filterCode.
        This trend
            is expected as the trigger components of the three applets
            are the same.
    
Figure~\ref{f:triggerserviceoverhead}~(b) shows the trigger service's
    network overhead.
This overhead is
    $2.5\times$ that of NoSec.
Also, as with the \cpu costs, the network traffic is independent
        of the filterCode.
    
\emparagraph{Action service overheads.}
Figure~\ref{fig:actionservicecpu-run} shows action service's \cpu and network
overhead for the various systems and filterCodes.
Similar to the trigger service overheads,
    the \cpu costs are dependent on the system used. 
The \cpu overhead of \sys relative to NoSec
    is due to the decryption and merging of actInp shares, and
        signature verification for the ``proofs'' from TEEs and the
            token-chain (\S\ref{s:design-active}, \S\ref{s:refreshtokens}).

The network overhead also increases with the security provided by the system,
and with the filterCode.
The variation across filterCodes is due to the fact that different filterCodes result
    in actInp of different lengths.
Finally, we notice that for NoSec 
    the action service-side network transfers (9.3~KB) are larger 
        than the platform-side network transfers (3.7~KB; 
Figure~\ref{f:servercpunetwork}~(b)).
This is because the action service also communicates with
        Gmail to authenticate
        and send emails.

\subsection{Dollar Cost}
\label{s:eval:dollarcost}
\Sys's added 
    overheads
    (\S\ref{s:eval:platformsideoverheads})
    increase dollar cost for the platform provider for running an applet. 
In this section, we estimate these costs
    by converting \sys's raw resource overheads
        into a dollar amount using
            a pricing model derived from machine and 
            data transfer prices 
            of Microsoft Azure (Appendix~\ref{a:pricingmodel}).
    This model charges \cpu at $\$0.198$/hour and network transfers at
             $\$0.087$/GB.

Figure~\ref{f:eval:dollarcost} shows the 
        resulting dollar costs 
    for
    the three types of
    filterCodes (passing around, string substitution, and custom code).
For the common filterCodes (passing around and string
    substitution), \sys increases cost by $3.7\times$ and $3.74\times$.
    However,
    for the custom code filterCode, \sys increases cost by $510.4\times$.

\subsection{Compatibility}
\label{s:eval:compatibility}
One of \sys's goals is to run
    any IFTTT applet. 
To study whether \sys can meet this requirement,
    we implemented,  using \sys, (i) 15 randomly selected IFTTT applets
        from the 50 most popular applets in
        the dataset of Mi et al.~\cite{mi2017empirical},
    and 
    (ii) 15 IFTTT applets whose applet IDs were randomly generated (Appendix~\ref{s:compatibilityappendix}).
        We found that 
            IFTTT applets can be implemented on \sys
                in a short time---it took us
                96 developer-hours to implement and test these applets.

\begin{figure}[t]
    \footnotesize
    \centering
    
    \begin{tabular}{
        @{}
            *{1}{>{\raggedright\arraybackslash}b{.1\textwidth}}  @{}   %
            *{1}{>{\raggedleft\arraybackslash}b{.07\textwidth}}         %
            *{1}{>{\raggedleft\arraybackslash}b{.07\textwidth}}         %
            *{1}{>{\raggedleft\arraybackslash}b{.09\textwidth}}         %
        @{}
        }
                  & \cpu time & Network transfers  & \$-cost relative to NoSec \\
        \midrule
        \multicolumn{2}{@{}l}{\textbf{passing around}} \\
        \multicolumn{1}{@{}l}{NoSec} &  17.6~ms & 3.6~KB & $1\times$ \\
        \multicolumn{1}{@{}l}{\Sys} & 62~ms  & 15.4~KB & $3.7\times$ \\
        \midrule
        \multicolumn{2}{@{}l}{\textbf{string substitution}} \\
        \multicolumn{1}{@{}l}{NoSec} &  17.6~ms & 3.7~KB & $1\times$ \\
        \multicolumn{1}{@{}l}{\Sys} & 62.9~ms    & 15.7~KB  & $3.74\times$ \\
        \midrule
        \multicolumn{2}{@{}l}{\textbf{custom filterCode}} \\
        \multicolumn{1}{@{}l}{NoSec} &  17.6~ms & 3.7~KB & $1\times$ \\
        \multicolumn{1}{@{}l}{\Sys} & 950.3~ms & 6966.6~KB & $510.4\times$ \\
        \bottomrule

    \end{tabular}
    \normalfont\selectfont
    \caption{Dollars spent  to run an applet on NoSec, \sys.}
    \label{f:eval:dollarcost}
\end{figure}

\section{Related work}
\label{s:relwork}

\emparagraph{Confidentiality and integrity risks in IoT middleware.}
Xu et
al.~\cite{xu2019privacy} study SmartThings, which is the Samsung
platform for managing Samsung smart devices, and report the movement of sensitive
information 
    from SmartThings to IFTTT.

Fernandes et al.~\cite{fernandes2016security} also report security issues with
SmartThings, in particular, 
    over-privilege of access tokens.        
For example, 
    a SmartThings application 
    may get a capability for locking and unlocking a Samsung device 
        even if it needs to only call lock.
Like SmartThings, IFTTT exhibits
over-privilege: 75\% 
of IFTTT services issue broad-scoped
OAuth tokens~\cite{fernandes2018decentralized, fernandes2017decoupled}, as
discussed earlier (\S\ref{s:background}).

\Sys is inspired by these works. It
    focuses on limiting the exposure of sensitive data
    at the platform and 
    and ensuring that the platform 
    does not misuse tokens (that is, always performs the precise action intended
        by the user).

\emparagraph{Reducing risks in trigger-action platforms.}
Almond~\cite{campagna2017almond} is a virtual assistant that connects trigger
and action services. Unlike IFTTT, it allows 
  the users to program applets using natural language (rather than mouse clicks
    that build automation programs).
Almond preserves confidentiality but makes strong trust assumptions: it runs
    inside a single trusted device.

Xu et al.~\cite{xu2019privacy} use a filter-and-fuzz approach
    to provide confidentiality.   
The filter part enables the trigger service to eliminate
    parameters that are either not needed by an applet or do not
    result in the firing of an action.
The fuzz part adds dummy trigger outputs to drown real outputs
    that cannot be filtered. 
Filter-and-fuzz approach has a limited 
    applicability as it assumes that the trigger service
    has knowledge
    of the actions. 
Indeed, Xu et al. apply their approach to Samsung's SmartThings
    where trigger and action services 
            are in the same domain.
Furthermore, 
    the approach
    adds high 
        \cpu and network overhead to transfer and process
    dummy events.

DTAP~\cite{fernandes2018decentralized, fernandes2017decoupled} eliminates
    over-privilege of OAuth tokens by using applet-specific access tokens.
In DTAP, a trigger service 
    signs trigOut, the platform processes
    the output in plaintext to generate actInp, and the action
    service verifies both the signature over trigOut and the
        correctness of actInp.
DTAP does not hide sensitive data from the platform. Besides, it sends
    the entire trigOut structure to the action service
    to verify signature and action-correctness.

\begin{figure}[t!]
\footnotesize
\centering

\usetkzobj{all}
\tikzset{%
    pics/sema/.style args={#1/#2/#3}{code={%
        \ifstrequal{#2}{0}{%
            \tkzDefPoint(0,0){O}
            \filldraw[fill=#1, draw=black] (0,0) circle (1.2mm);
        }{%
            \tkzDefPoint(0,0){O}
            \tkzDrawSector[R,fill=#1](O,1.2mm)(90,90-#2)
            \tkzDrawSector[R,fill=#3](O,1.2mm)(90-#2,90-360)
    }
    }},
}

\newcommand{\tablecolwidth}{1ex}
\DeclareRobustCommand{\zerofull}{\tikz\pic{sema=white/0/};}
\DeclareRobustCommand{\lowfull}{\tikz\pic{sema=white/90/black};}
\DeclareRobustCommand{\mediumfull}{\tikz\pic{sema=white/180/black};}
\DeclareRobustCommand{\highfull}{\tikz\pic{sema=white/270/black};}
\DeclareRobustCommand{\full}{\tikz\pic{sema=black/0/};}

\begin{tabular}{
@{}
*{1}{>{\raggedright\arraybackslash}b{.115\textwidth}} @{} 
*{1}{>{\centering\arraybackslash}b{.08\textwidth}}   @{\hspace{\tablecolwidth}}%
*{1}{>{\centering\arraybackslash}b{.05\textwidth}}   @{\hspace{\tablecolwidth}}%
*{1}{>{\centering\arraybackslash}b{.08\textwidth}}  @{\hspace{\tablecolwidth}}   %
*{1}{>{\centering\arraybackslash}b{.058\textwidth}} @{\hspace{\tablecolwidth}}  %
*{1}{>{\centering\arraybackslash}b{.057\textwidth}} @{\hspace{\tablecolwidth}}  %
@{}
}

 & Functionality & Conf. & Data Min. & Integrity & Efficiency \\ 

\midrule

DTAP~\cite{fernandes2018decentralized, fernandes2017decoupled} & \full & \zerofull & \zerofull & \full & \full \\
Filter-Fuzz~\cite{xu2019privacy} & \full & \mediumfull  & \zerofull & \zerofull & \mediumfull \\
\Sys (\S\ref{s:overview}--\S\ref{s:refresh-tokens})  & \full & \highfull & \full & \full & \highfull \\

\bottomrule

\end{tabular}
\normalfont\selectfont
\caption{%
Comparison of \sys to prior work 
    for trigger-action
platforms.
An empty circle ($\zerofull{}$) denotes that the system does not address the
particular concern,
while a full circle 
($\full{}$) denotes that the system performs close to the ideal. 
The other circles
    fall in between in an increasing
order of superiority. 
}

\label{fig:relwork}
\label{f:relwork}
\end{figure}

\Sys also targets security improvements 
    for a trigger-action platform.
However, relative to prior work, it provides
    both confidentiality and integrity guarantees (Figure~\ref{fig:relwork}). 
Furthermore, in line with the principle of least
privilege~\cite{saltzer1975protection, denning1976fault} (also called data
minimization in modern privacy regulation~\cite{gdpr}), 
    \sys ensures that the services 
    learn only the information that is input to the trigger and action APIs.  
    
\emparagraph{Reducing risk in other types of IoT middleware.}
Bolt~\cite{gupta2014bolt} targets applications for smart homes (e.g.,
finding a lost
    pet in a neighborhood). It
    stores data on an untrusted cloud provider but runs
    applications on trusted in-home devices.

BeeKeeper~\cite{zhou2018beekeeper} targets a streaming scenario in which an
    IoT device streams data to a blockchain-based ``leader'' 
    who runs a query on it.
BeeKeeper uses secret-sharing and secure multi-party
computation~\cite{goldreich1987play} to provide 
    confidentiality, and threshold-cryptography to provide 
    integrity. However, 
                BeeKeeper does not target trigger-action scenarios; instead, it
        computes       
        quadratic polynomials (mean, variance, and standard-deviation).        
Jayaraman et al.~\cite{jayaraman2017privacy}
    target the same scenario as BeeKeeper. However, 
    in contrast to BeeKeeper, they use
        Paillier encryption and do not provide integrity.

\emparagraph{General-purpose crypto primitives for security.}
Secure computation~\cite{yao1982protocols,goldreich1987play} and homomorphic
encryption~\cite{gentry2009fully,rivest1978data} 
    are vibrant areas of research~\cite{kolesnikov2008improved,
kolesnikov2014flexor, zahur2015two, malkhi2004fairplay, huang11faster,
holzer2012secure, songhori2015tinygarble, chillotti2016faster,
chillotti2020tfhe}). These
general-purpose primitives support arbitrary computation but are 
    expensive.
\Sys uses Yao's two-party protocol 
but only for a small fraction of applets (\S\ref{s:design-passive}).

Zero-knowledge proof systems
    have also seen tremendous development~\cite{setty2018proving,
backes15adsnark, walfish15verifying, bensasson14succinct, braun13verifying,
parno13pinocchio, chiesa15cluster, costello15geppetto}. 
However, the party generating the proof 
    incurs a large \cpu time.

\sys instead uses a mix of digital signatures and TEEs for
    proving execution integrity at low overhead (\S\ref{s:design-active}).

\section{Conclusion}
\label{s:summary}
\label{s:discussion}
\label{s:conclusion}

Trigger-action platforms like IFTTT have gained traction due to their convenience
and connectivity. 
However, these cloud-hosted platforms present confidentiality
and integrity risks.    
This paper asked the question, can we build a fortified alternative to IFTTT
    at a low resource cost, and found the results to be encouraging.
In particular, one can gain significantly on security while 
    restricting resource (\cpu, network) overhead
    to 4.3$\times$ for all but a small 
        fraction of programs. \Sys's enabler is a new secure
        computation protocol that is tailored for common computation on IFTTT
        and that distributes trust over heterogenous
        trusted-hardware machines from different vendors.
By demonstrating these promising results, \sys
    provides a benefit to
    both users and platform providers---they can enjoy trigger-action automation with
    significant improvement in data confidentiality and integrity.

\appendix

\section{Pricing model}
\label{a:pricingmodel}

To get the dollar-cost of running an applet, 
    we take the \cpu and network cost from our benchmarks and add them linearly
(\S\ref{s:eval:dollarcost}). 
    That is, if $C$ is the hourly \cpu cost and $D$ is the price for
transferring one GB of data over the network,
            the dollar cost to run each applet is of the form $\alpha C + \beta
D$, where
    $\alpha$ is the \cpu time required to run the applet and $\beta$ is the
amount of data transferred to or from the \sys platform during execution. 
    Our goal in this appendix is to estimate the values of $C$ and $D$.

The network cost is taken from the
``outbound data transfer'' costs for Microsoft Azure which gives $D = 0.087$
(dollars per GB)~\cite{azurebandwidthpricingmodel}.
Meanwhile, estimating \cpu cost is not straightforward. 
Machine prices on cloud providers are set using 
     several factors
        including maintenance and running expenses, 
            market demand,  and 
        peripheral costs such as DDoS protection and
        load-balancing.
Therefore, we make a simplifying assumption that the 
  machine cost on cloud providers is dominated by
        the cost of \cpu.
We take the hourly cost of a DC4s\_v2 SGX machine on Microsoft
Azure~\cite{azuremachinepricingmodel} that we use in our experiments 
    and divide by 4, which is the number of cores on
this machine. This method
gives $C = 0.198$ (dollars/hour). 

\section{Compatibility study applets}
\label{s:compatibilityappendix}

\begin{myitemize}

    \item The following 15 applets were randomly selected from the set of the 50
most popular applets from the dataset of Mi et al.~\cite{mi2017empirical}:
    \begin{myenumerate}
        \item ``Get a daily 6:00 AM email with the weather report'' (\url{https://ifttt.com/applets/Zv56ZXwR})
        \item ``Update your Android wallpaper with NASA's image of the day'' (\url{https://ifttt.com/applets/yNvHX9VQ})
        \item ``Sync all your new iOS Contacts to a Google Spreadsheet'' (\url{https://ifttt.com/applets/dycqQ5A6})
        \item ``Automatically change your Twitter profile pic when you update your Facebook photo'' (\url{https://ifttt.com/applets/qFZqXrvs})
        \item ``Automatically back up your new iOS photos to Google Drive'' (\url{https://ifttt.com/applets/90254p})
        \item ``Send a text message to someone with your Android and Google Home'' (\url{https://ifttt.com/applets/fNdGJfwy})
        \item ``Sync your Amazon Alexa to-dos with your reminder'' (\url{https://ifttt.com/applets/ieCE52WK})
        \item ``Track your work hours in Google Calendar'' (\url{https://ifttt.com/applets/sFk2WC4r})
        \item ``Quickly create events in Google Calendar'' (\url{https://ifttt.com/applets/192007p})
        \item ``Press a button to track work hours in Google Drive'' (\url{https://ifttt.com/applets/227069p})
        \item ``Back up new iOS photos you take to Dropbox'' (\url{https://ifttt.com/applets/103376p})
        \item ``Backup your new Instagram photos to Dropbox'' (\url{https://ifttt.com/applets/25679p})
        \item ``Post your Instagram photos to Tumblr'' (\url{https://ifttt.com/applets/131p})
        \item ``Post your Instagram photos as native Twitter photos when \#twitter is in the caption'' (\url{https://ifttt.com/applets/68915p})
        \item ``Tell Alexa to email you your shopping list'' (\url{https://ifttt.com/applets/284243p})
    \end{myenumerate}

    \item The following 15 applets were randomly selected from a set of 78 randomly generated applet IDs:
    \begin{myenumerate}
        \item ``bloggin' via Instagram: STEP 2 Flickr -\textgreater Tumblr'' (\url{https://ifttt.com/applets/7p})
        \item ``End Harmony Activity using MESH'' (\url{https://ifttt.com/applets/6163})
        \item ``Receive a notification when occupancy is detected'' (\url{https://ifttt.com/applets/3820})
        \item ``Turn on Wemo Smart Plug using MESH'' (\url{https://ifttt.com/applets/6190})
        \item ``bloggin' via Instagram: STEP 1 Instagram -\textgreater Flickr'' (\url{https://ifttt.com/applets/6p})
        \item ``Tweet your newest uploaded Flickr pictures.'' (\url{https://ifttt.com/applets/29p})
        \item ``Carry important file everywhere!'' (\url{https://ifttt.com/applets/86p})
        \item ``Record each time you connect to your TP-Link network in a Google Spreadsheet'' (\url{https://ifttt.com/applets/6997})
        \item ``Test ifttt 1'' (\url{https://ifttt.com/applets/09018})
        \item ``Switch on a socket with Amazon Alexa'' (\url{https://ifttt.com/applets/6569})
        \item ``Flickr favorites to Evernote'' (\url{https://ifttt.com/applets/33p})
        \item ``Get an email when motion is detected at your front door'' (\url{https://ifttt.com/applets/3726})
        \item ``Sing a love song to IFTTT Voicemail, send it straight to my sweetheart'' (\url{https://ifttt.com/applets/11p})
        \item ``Daily Forecast'' (\url{https://ifttt.com/applets/2950p})
        \item ``Post your new Instagram videos with a specific hashtag to a Telegram chat'' (\url{https://ifttt.com/applets/4657})
    \end{myenumerate}

\end{myitemize}

\section{Security proofs}
\label{s:proofAppendix}

We first introduce definitions for the 
    basic cryptographic primitives such as secret-sharing, encryption, and signatures
that \sys's protocol builds on. We then present the proofs for the passive and
    active security settings (\S\ref{s:security-definitions}).

\subsection{Basic cryptographic primitives}

\emparagraph{Secret-sharing.}
\sys uses a 2-out-of-2 XOR secret-sharing scheme, where both
the shares are required to reconstruct the secret. This scheme consists
of two algorithms.

\begin{myitemize2}
\item $\share(s \in \{0,1\}^{\ell}) \rightarrow (k, k')$ 
    takes as input a $\ell$-bit string and generate two secret-shares
    of the string: $k$ and $k'$. Internally, the procedure 
    first samples $k \in_R \{0, 1\}^{\ell}$ and 
        then computes $k' \gets k \oplus s$.

\item $\reconstruct(k, k')$ outputs $s \gets k \oplus k'$.
\end{myitemize2}

\noindent Correctness of the scheme requires the shares produced to be reconstructable. 
Security requires that an adversary cannot distinguish between the 
    shares of two different input strings of the same length.

\emparagraph{Encryption scheme.}
\sys relies on an encryption scheme that is IND-CPA and IND-CCA2 secure. It consists
of two algorithms.

\begin{myitemize2}
\item $\enc(pk, m) \rightarrow C^{m}$ takes in the public key, and the plaintext
message and outputs a ciphertext.

\item $\dec(sk, C^{m}) \rightarrow m$ takes in the secret key and  a ciphertext
and outputs the plaintext message.
\end{myitemize2}

\noindent Correctness requires that ciphertexts produced from the scheme can be decrypted to 
obtain back the original plaintext. Security requires 
    that the ciphertexts produced for two different plaintext messages are
indistinguishable. \Sys's implementation uses
    the elliptic-curve based ECIES encryption scheme (\S\ref{s:eval}).

\emparagraph{Signature scheme.}
\sys relies on a standard signature scheme which consists of two algorithms.

\begin{myitemize2}
\item $\sign(sk, m) \rightarrow \sigma_m$ takes in a secret key and a message
and outputs a signature.

\item $\verify(pk, m, \sigma_m) \rightarrow \{\text{True}, \perp\}$ takes in the 
public key, the message, and the signature, and returns either True or $\perp$ to 
indicate whether the signature on the message is valid or not.
\end{myitemize2}

\noindent Correctness requires that a signature produced from the scheme can be successfully 
verified by the verification algorithm. Security requires the 
unforgeability of a valid signature by an adversary, that is, 
    an adversary cannot (without the secret key) produce a signature that
outputs True
    from the verification algorithm.

We now move to the proofs for \sys's protocol.

\subsection{Proof for passive security}
\label{a:passive-security-proof}

Our goal here is to prove that the protocol described in
Section~\ref{s:design-passive} (Figure~\ref{f:forttt-passive-protocol}) 
    meets the passive security definition in
    Section~\ref{s:security-definitions}.

We prove the security of the protocol using the simulation proof technique.
Essentially, in this technique, there is an ideal world 
adversary called the simulator which resides in the ideal world (which is secure by definition). The real world adversary interacts with the 
simulator, and the simulator simulates the messages output by the honest parties in the protocol. The simulator has access to the 
        signing keys of the honest parties. 
The protocol is said to be secure if such a simulator exists and the adversary cannot distinguish between the real execution 
of the protocol and the simulated execution. 

Suppose there exists a PPT adversary $\adversary$ that compromises the general purpose machine $M_b$ for some $b \in \{0, 1\}$. 
The honest parties in the passive setting are trigger service $TS$, action
service $AS$, and the general-purpose machine $M_{1-b}$.
We then construct a PPT simulator $\simr$ as follows: 

\begin{myenumerate4}
\item $\simr$ chooses a uniform random tape for the corrupted party $M_b$. 
        $\simr$ simulates the honest parties $TS, AS, M_{1-b}$.

\item \textbf{In Setup phase:}
      \begin{myenumerate}
      \item $\simr$ sends encryptions of zeros as access tokens and trigInp. That is, 
      $C^{at\mhyphen TS}_{\simr} \gets \enc(pk_{TS}, 0^{|at\mhyphen TS|})$,
$C^{at\mhyphen AS}_{\simr} \gets \enc(pk_{AS}, 0^{|at\mhyphen AS|})$,
       and $C^{trigInp}_{\simr} \gets \enc(pk_{TS}, 0^{|trigInp|})$.

      \item $\simr$ replaces each $sh_b^{(t_i)}$ not of the form \{\{key-name\}\} in each $sh_b^{(actInp)}[k]$ with Share($0^{|t_i|}$)[$b$].

      \end{myenumerate}
\item \textbf{In trigger-polling phase:}
        $\simr$ replaces each $sh_b^{(t_i)}$ in each $sh_b^{(trigOut)}[k]$ with Share($0^{|t_i|}$)[$b$].
      
\item \textbf{In action-generation phase:}
        depending on the filterCode, $\simr$ runs the simulator of Yao's
protocol or the custom string substitution protocol (described shortly below).
      In either case, the simulator is programmed to output shares of the action input parameter $sh_b^{(actInp)}$ chosen uniformly at random.
      
\item \textbf{In action-execution phase:} $\simr$ does nothing.
\end{myenumerate4}

\noindent We now show that the 
    distribution of the messages in the real protocol and the outputs of the
simulator are 
    computationally indistinguishable using 
the standard hybrid technique.
This technique 
refers to the strategy of proving that two
tuples of distributions $(D_1,\ldots,D_n)$ and $(D'_1,\ldots,D'_n)$ are
computationally indistinguishable, if all the $D_i$s (respectively, $D'_i$s) are
independently generated, and $D_i$ is computationally indistinguishable from
$D'_i$. The strategy is to consider a sequence of hybrid distributions
$(D_1,\ldots,D_{i-1},D'_i,\ldots,D'_n)$ and proving every consecutive tuples of
distributions are computationally indistinguishable.

\begin{myenumerate4}
\item $Hyb_0$: This corresponds to the real world distribution where trigger service $TS$, action service $AS$ and machine $M_{1-b}$
execute the protocol as described in 
Section~\ref{s:design-passive}.

\item $Hyb_1$: In this hybrid, we change the ciphertexts of tokens and trigInp sent to $TS$, and ciphertext of tokens sent to $AS$.
More specifically, we replace the encryption of tokens and trigInp with a bit strings chosen uniformly at random. 
$Hyb_0 \approx_c Hyb_1$ due to the semantic security of the encryption scheme. 

\item $Hyb_2$: In this hybrid, we change the shares of actInp and trigOut.
More specifically, each $sh_b^{(t_i)}$ not of the form \{\{key-name\}\} in each $sh_b^{(actInp)}[k]$ and $sh_b^{(trigOut)}[k]$ is replaced
with bit strings chosen uniformly at random. $Hyb_1 \approx_c Hyb_2$ due to the security of the secret sharing scheme.

\item $Hyb_3$: In this hybrid, the simulators for either Yao's protocol or custom string substitution protocol is invoked which
outputs honest shares of action input parameter $sh_b^{(actInp)}$. $Hyb_2 \approx_c Hyb_3$ due to the simulation security of 
Yao's and custom string substitution protocols (we prove the latter below).

\item $Hyb_4$: In this hybrid, we replace $sh_b^{(actInp)}$ output from either Yao's protocol or custom string substitution protocol 
with bit strings chosen uniformly at random. $Hyb_3 \approx_c Hyb_4$ due to the
security of secret-sharing.

\item $Hyb_5$: This corresponds to the output distribution of $\simr$. Hybrids $Hyb_4$ and $Hyb_5$ are identically distributed.
\end{myenumerate4}

\emparagraph{Proof of security for custom string substitution.}

\begin{definition}
The custom string substitution protocol described in Figure~\ref{f:forttt-action-field-generation-code} is said to 
be $\mathcal{L}\mhyphen$secure if for any honest-but-curious (passive) probabilistic polynomial time (PPT) adversary $\adversary$ corrupting
$M_b$ for some $b \in \{0,1\}$, for a leakage function $\mathcal{L}$, for every large enough security parameter $\lambda$, 
there exists a PPT simulator $\simr$ with access to the leakage function, such that 
        the output distribution of the simulator is computationally
            indistinguishable from the adversary's view in the real protocol.

\end{definition}

\noindent The simulator for string\_sub is constructed as follows:
\begin{myenumerate4}

\item $\simr$ replaces each $sh_b^{(t_i)}$ in each $sh_b^{(trigOut)}[k]$ with
Share($0^{|t_i|}$)[$b$].

\item Similarly, $\simr$ replaces each $sh_b^{(t_i)}$ not of the form
\{\{key-name\}\} in each $sh_b^{(actInp)}[k]$ with Share($0^{|t_i|}$)[$b$].

\item $\simr$ replaces each $sh_b^{(t_i)}$ in each $sh_b^{(actInp)}[k]$ output from GenerateAI with Share($0^{|t_i|}$)[$b$] and outputs $sh_b^{(actInp)}$.

\end{myenumerate4}

\noindent We now show that the 
    distribution of the messages in the real protocol and the outputs of the
simulator are 
    computationally indistinguishable using 
the hybrid technique.

\begin{myenumerate4}
\item $Hyb_0$: This hybrid corresponds to the real world distribution of the execution of string\_sub function by machine $M_b$. That is,
\{string\_sub($sh_b^{(trigOut)}, sh_b^{(actInp)}$)\}.

\item $Hyb_1$: In this hybrid, we change the share of trigOut to secret shares of zero.
More specifically, each $sh_b^{(t_i)}$ in each $sh_b^{(trigOut)}[k]$ is replaced
with Share($0^{|t_i|}$)[$b$].

\item $Hyb_2$: In this hybrid, we change the share of actInp input parameter to secret shares of zero.
More specifically, each $sh_b^{(t_i)}$ not of the form \{\{key-name\}\} in each $sh_b^{(actInp)}[k]$ is replaced
with Share($0^{|t_i|}$)[$b$]. $Hyb_0 \approx_c Hyb_1 \approx_c Hyb_2$ due to the semantic security of secret sharing scheme.

\item $Hyb_3$: This corresponds to the output distribution of $\simr$. 
Hybrids $Hyb_2$ and $Hyb_3$ are identically distributed. 

\end{myenumerate4}

\subsection{Proof for active security}
\label{a:active-security-proof}

Our goal here is to prove that the protocol described in
Section~\ref{s:design-active} (Figure~\ref{f:forttt-active-protocol}) and
Section~\ref{s:refresh-tokens}
    meets the active security definition in
    Section~\ref{s:security-definitions}.
We first formalize the notion of token-chains (\S\ref{s:refresh-tokens}) and
then present the proof.

\begin{definition}[Token-chains]
Let $at\mhyphen TS_{\ell}$ and $rt\mhyphen TS_{\ell}$ denote the access and
refresh token for trigger service $TS$ in epoch $\ell$, and let 
$at\mhyphen AS_{\ell}$ and $rt\mhyphen AS_{\ell}$ denote the access and refresh token 
    for action service $AS$ for epoch $\ell$.
We define $chain_{\ell}^{TS}$ equal to $\enc(pk_{TS}, at\text{-}TS_0 ||
at\text{-}TS_{\ell})$ as the token chain for the trigger service TS for epoch $\ell$, and 
$\sigma_{\ell}^{TS} \leftarrow \sign(sk_{TS}, chain_{\ell}^{TS})$ as the
signature over the chain. 
Similarly, we define $chain_{\ell}^{AS}$ and $\sigma_{\ell}^{AS}$. \Sys's
protocol for active security adds
    $chain_{\ell}^{TS}, \sigma_{\ell}^{TS}, chain_{\ell}^{AS}$, and
$\sigma_{\ell}^{AS}$ to
$\textrm{App}_b$.

\end{definition}

Suppose there exists a PPT adversary $\adversary$ that compromises the general
purpose machine $M_b$ for $b \in \{0, 1\}$ and TEE machines $T_b^{(i)}$ and
$T_b^{(j)}$ for some $ i , j \in \{0, 1, 2\}$. 
Then the honest parties in the active setting are trigger service $TS$, action service $AS$, TEEs 
$T_b^{(k)}, T_{1-b}^{(i)}, T_{1-b}^{(j)}, T_{1-b}^{(k)}$, and the general purpose machine $M_{1-b}$.
Let $k$ be such that $k \neq i$ and $k \neq j$. We then construct a PPT
simulator $\simr$ simulating the behavior of the honest parties as follows.

\begin{myenumerate4}

\item \textbf{In setup phase:} 
      \begin{myenumerate4}
      \item $\simr$ sends encryptions of zeros as access tokens and trigInp, that is, 
      $C^{at\mhyphen TS}_{\simr} \gets \enc(pk_{TS}, 0^{|at\mhyphen TS|})$,
$C^{at\mhyphen AS}_{\simr} \gets \enc(pk_{AS}, 0^{|at\mhyphen AS|})$,
       and $C^{trigInp}_{\simr} \gets \enc(pk_{TS}, 0^{|trigInp|})$.

      \item $\simr$ replaces each $sh_b^{(t_i)}$ not of the form
\{\{key-name\}\} in each $sh_b^{(actInp)}[key]$ with Share($0^{|t_i|}$)[$b$].

      \item $\simr$ sends encryptions of zeros as the token chains, that is,
      $chain_{{\ell}_{\simr}}^{TS} \leftarrow \enc(pk_{TS}, 0^{|chain_{\ell}^{TS}|})$,
$chain_{{\ell}_{\simr}}^{AS} \leftarrow \enc(pk_{TS}, 0^{|chain_{\ell}^{AS}|})$.

      \item $\simr$ generates simulated signatures $\sigma_{\ell}^{TS},
\sigma_{\ell}^{AS}, \sigma_T$ for the uniform random ciphertexts
$chain_{{\ell}_{\simr}}^{TS}, chain_{{\ell}_{\simr}}^{AS}$, 
      and simulated $T$, respectively.
      \end{myenumerate4}

\item \textbf{In trigger-polling phase:} 
      \begin{enumerate}
      \item $\simr$ asks $\adversary$ for its inputs to the trigger service $TS$: $T,
            \sigma_T, chain_{\ell}^{TS}, \sigma_{\ell}^{TS}$.
            $\simr$ additionally aborts the protocol if signature verification
of $\sigma_T$ and $\sigma_{\ell}^{TS}$ fails.

      \item $\simr$ simulates trigger output $tout_b$ by generating simulated
                shares of $sh_b^{(trigOut)}$: replacing each $sh_b^{(t_i)}$ in
$sh_b^{(trigOut)}[key]$ with Share($0^{|t_i|}$)[$b$] and
      generating simulated signature $\sigma_{tout_b}$ on $tout_b$.

      \item $\simr$ then sends $tout_b$ and $\sigma_{tout_b}$ to $\adversary$.
      \end{enumerate}

\item \textbf{In action-generation phase:} 
\begin{enumerate}

\item $\simr$ asks $\adversary$ for its input $tout_b$ to the non-compromised
TEE $T_b^{(k)}$. $\simr$ runs the checks
described in step~\ref{e:matchTID} of Figure~\ref{fig:forttt-active-protocol} and aborts the protocol if verification fails.

\item $\simr$ simulates the generated action input parameter shares $sh_b^{(actInp_k)}$
of TEE $T_b^{(k)}$ by sampling each $sh_b^{(actInp_k)}[key]$ for $(key, v)$ in
$sh_b^{(actInp_k)}$ uniformly at random. $\simr$ also generates simulated
$\text{proof} \mhyphen T_b^{(k)}$ on 
the simulated $ain_b^{(k)}$.

\item $\simr$ then depending on the filterCode, invokes the simulator for
maliciously secure Yao's protocol (this is Yao's protocol secure against
honest-but-curious adversaries but with
            the underlying oblivious 
        transfer sub-protocol replaced by a variant of oblivious transfer secure against malicious
adversaries~\cite{huang2012quid, keller2015actively})
        and generates simulated action input shares 
    $sh_b^{(actInp_i)}$, $sh_b^{(actInp_j)}$, and proofs $\text{proof} \mhyphen T_b^{(i)}$ and $\text{proof} \mhyphen T_b^{(j)}$ 
as described in the above step.

\item $\simr$ sends $ain_b^{(i)}, ain_b^{(j)}, ain_b^{(k)}, \text{proof} \mhyphen T_b^{(i)}, \text{proof} \mhyphen T_b^{(j)}$ 
and $\text{proof} \mhyphen T_b^{(k)}$ to $M_b$.
\end{enumerate}

\item \textbf{In action-execution phase:}
\begin{myenumerate}
\item $\simr$ asks $\adversary$ for its inputs to the action service $AS$: 
$ainp_b, \text{proof} \mhyphen T_b^{(i)}, \text{proof} \mhyphen T_b^{(j)},
\text{proof} \mhyphen T_b^{(k)}, chain_{\ell}^{AS}$,
and $\sigma_{\ell}^{AS}$.
\item $\simr$ runs step~\ref{e:checkmatch} of
Figure~\ref{fig:forttt-active-protocol} and checks $RID$ as mentioned in 
step~\ref{e:executeaction}  of Figure~\ref{fig:forttt-active-protocol}.
\end{myenumerate}

\item \textbf{In token-refresh phase:} 
\begin{enumerate}
\item $\simr$ asks $\adversary$ for its inputs to the $TS$:
$chain_{\ell}^{TS}$ and $rt\mhyphen TS$. $\simr$ checks
the signature $\sigma_{\ell}^{TS}$ on $chain_{\ell}^{TS}$ and aborts the
protocol if the verification fails.  

\item $\simr$ generates simulated chain
$chain_{\ell+1}^{TS} \gets \enc(pk_{TS}, 0^{|chain_{\ell}^{TS}|})$, and 
signature $\sigma_{\ell+1}^{TS} \leftarrow \sign(sk_{TS}, chain_{\ell+1}^{TS})$. $\simr$
sends $chain_{\ell+1}^{TS}$ and $\sigma_{\ell+1}^{TS}$ to $\adversary$.
\end{enumerate}

$\simr$ behaves symmetrically if the token-refresh request is for $AS$.
\end{myenumerate4}

\noindent 
We now show that the distribution of the messages in the real
protocol and the outputs of the simulator are computationally indistinguishable
using the hybrid technique.

\begin{myenumerate4}

\item $Hyb_0$: This hybrid corresponds to the real world distribution where
trigger service $TS$, action service $AS$, machine $M_{1-b}$, 
    and TEE machines $T_0^{(k)}$ and $T_1^{(k)}$ execute the protocol described in Figure~\ref{fig:forttt-active-protocol}.

\item $Hyb_1$: In this hybrid, we change the ciphertexts of tokens and trigInp sent to $TS$, and ciphertext of tokens sent to $AS$. 
More specifically, we replace the encryption of tokens and trigInp with bit strings chosen uniformly at random. The signatures 
$\sigma_T, \sigma_{\ell}^{TS}$ and $\sigma_{\ell}^{AS}$ are also updated accordingly.
$Hyb_1 \approx_c Hyb_0$ due to the semantic security of the encryption scheme.

\item $Hyb_2$: In this hybrid, the shares of trigOut in $tout_b$ obtained as the output of trigger-polling phase $sh_b^{(trigOut)}$ is replaced
with a bit string sampled uniformly at random. The signature $\sigma_{tout_b}$ is updated according to the simulated value of $tout_b$. 
$Hyb_2 \approx_c Hyb_1$ due to the security of secret-sharing.

\item $Hyb_3$: This hybrid is same as $Hyb_2$ but $\simr$ verifies $\sigma_T,
\sigma_{\ell}^{TS}, \sigma_{\ell}^{AS}$ in the trigger-polling phase and aborts if the
verification fails.  $Hyb_3$ is identically distributed to $Hyb_2$.

\item $Hyb_4$: In this hybrid, the action generation output of $T_b^{(k)}$, $sh_b^{(actInp_k)}$ is replaced with a bit string sampled
uniformly at random. The signature $\text{proof}\mhyphen T_b^{(k)}$ is updated
according to the simulated value of $ain_b^{(k)}$.
$Hyb_4 \approx_c Hyb_3$ due to the security of the secret-sharing scheme.

\item $Hyb_5$: In this hybrid, the simulator of Yao's protocol that outputs
honest shares $sh_b^{(actInp_{i})}, sh_b^{(actInp_{j})}$ is invoked. 
$Hyb_5 \approx_c Hyb_4$ due to the simulation security of Yao's protocol.

\item $Hyb_6$: In this hybrid, the shares $sh_b^{(actInp_{i})},
sh_b^{(actInp_{j})}$ output from the simulator of Yao is replaced with bit 
strings chosen uniformly at random. The signatures $\text{proof}\mhyphen T_b^{(i)}$ and $\text{proof}\mhyphen T_b^{(j)}$ are updated according to the 
simulated value of $ain_b^{(i)}$, $ain_b^{(j)}$. $Hyb_6 \approx_c Hyb_5$ due to the security of the secret-sharing scheme.

\item $Hyb_7$: In this hybrid, $\simr$ additionally runs signature checks mentioned in step~\ref{e:matchTID} of Figure~\ref{fig:forttt-active-protocol} and
aborts if verification fails. $Hyb_7$ is identically distributed to $Hyb_6$.

\item $Hyb_8$: In this hybrid, $\simr$ additionally runs checks mentioned in step~\ref{e:checkmatch} of Figure~\ref{fig:forttt-active-protocol} and aborts if
verification fails. $Hyb_8$ is identically distributed to $Hyb_7$.

\item $Hyb_9$: This hybrid corresponds to the output distribution of $\simr$ in the ideal execution. $Hyb_9$ and $Hyb_8$ are identically distributed.
\end{myenumerate4}

\section*{Acknowledgments}
We thank
Prabhanjan Ananth,
Gareth George,
Muqsit Nawaz,
Srinath Setty,
Yang Wang, and
Rich Wolski for 
    feedback and comments that helped improve 
    this draft.

\frenchspacing

\begin{flushleft}
\setlength{\parskip}{0pt}
\setlength{\itemsep}{0pt}
\bibliographystyle{abbrv}
\bibliography{conferences-long-with-abbr2,forttt}
\end{flushleft}
\label{p:last}
\end{document}